\documentclass{pasj00}

\begin{document}
\SetRunningHead{Author(s) in page-head}{Running Head}
\Received{2007/10/20}
\Accepted{2008/10/07}

\title{Uniting the Quiescent Emission and Burst Spectra of Magnetar Candidates}

\author{Yujin E. Nakagawa,\altaffilmark{1,2}
Atsumasa Yoshida,\altaffilmark{1}
Kazutaka Yamaoka,\altaffilmark{1}
Noriaki Shibazaki,\altaffilmark{3}
}
\altaffiltext{1}{Graduate School of Science and Engineering, Aoyama Gakuin University, 5-10-1 Fuchinobe, \\Sagamihara, Kanagawa 229-8558}
\email{yujin@crab.riken.jp}
\altaffiltext{2}{Institute of Physical and Chemical Research (RIKEN), 2-1 Hirosawa, Wako, Saitama 351-0198}
\altaffiltext{3}{Department of Physics, Rikkyo University, Nishi-Ikebukuro, Toshima-ku, Tokyo 171-8501}



\maketitle

\begin{abstract}
Spectral studies of quiescent emission and
bursts of magnetar candidates
using XMM-Newton, {\it Chandra} and {\it Swift} data are presented.
Spectra of both the quiescent emission and the bursts for most magnetar
candidates are reproduced by a photoelectrically absorbed two blackbody function (2BB).
There is a strong correlation between lower and higher temperatures of 2BB
($kT_{\mathrm{LT}}$ and $kT_{\mathrm{HT}}$) for the magnetar candidates of which
the spectra are well reproduced by 2BB.
In addition, a square of radius for $kT_{\mathrm{LT}}$ ($R_{\mathrm{LT}}^2$) is well
correlated with a square of radius for $kT_{\mathrm{HT}}$ ($R_{\mathrm{HT}}^2$).
A ratio $kT_{\mathrm LT}/kT_{\mathrm HT} \approx 0.4$
is nearly constant irrespective of
objects and/or emission types (i.e., the quiescent emission
and the bursts).
This would imply a common emission mechanism among the magnetar candidates.
The relation between the quiescent emission and the bursts
might be analogous to a relation between microflares and
solar flares of the sun.
Three AXPs (4U\,0142$+$614, 1RXS\,J170849.0$-$400910 and 1E\,2259$+$586)
seem to have an excess above $\sim$\,7\,keV which
well agrees
with a non-thermal hard component discovered by INTEGRAL.
\end{abstract}

\section{Introduction}
Among peculiar celestial objects in the universe, a dense highly
magnetized neutron star ($\rho \sim 10^{14}$\,g\,cm$^{-3}$ and $B \sim 10^{15}$\,G),
so-called ``magnetar'' \citep{duncan1992, paczynski1992, thompson1995, thompson1996},
would be one of the most exotic objects.
Soft gamma repeaters (SGRs) and anomalous X-ray pulsars (AXPs) are
well known as magnetar candidates.
An apparent difference between the SGRs and the AXPs would be
considered from their first detections.
The SGRs were discovered as sporadically bursting objects,
while the AXPs were regarded as peculiar pulsars with
long spin periods.
However, current observations unveil a lot of similarities between
these objects.
They have, for instance, long spin periods
($P \sim $ 5-12\,s) with spindown rates
of $\dot{P} \sim 10^{-10}$-$10^{-13}$\,s\,s$^{-1}$,
no signature of a companion star,
a distribution around the galactic plane
(two magnetar candidates are in
other galaxies), quiescent soft X-ray emission.
Several of these objects have non-thermal hard ($>20$\,keV)
components, some are associated with supernova remnants (SNRs),
and bursting activity is not confined to the SGRs but is observed
in the AXPs as well.
Considering these similarities, 
the SGRs and the AXPs should be classified into a common class of
objects.

So far, five SGRs (0501$+$4516, 0526$-$66, 1627$-$41, 1806$-$20 and 1900$+$14) are
known \citep{woods2006, barthelmy2008} as well as three candidates, SGR\,1801$-$23 \citep{cline2000},
SGR\,1808$-$20 \citep{lamb2003} and SGR/GRB\,050925.
SGR/GRB\,050925 was regarded as a gamma-ray burst (GRB) when
first detected, but soon after was recognized
as a new SGR \citep{holland2005}.
On the other hand, ten AXPs (1E\,2259$+$586, 1E\,1048.1$-$5937,
4U\,0142$+$614, 1RXS\,J170849.0$-$400910, 1E\,1841$-$045, XTE\,J1810$-$197,
AX\,J1845$-$0258, CXOU\,J010043.1$-$721134, CXOU\,J164710.2$-$455216 and 1E\,1547.0$-$5408)
are known to date \citep{woods2006, dib2008}
with one AXP candidate,
AXP\,CXOU\,J160103.1$-$513353 \citep{park2006}.
A short burst from AXP\,CXOU\,J164710.2$-$455216 was
detected by {\it Swift} BAT \citep{krimm2006} at 01:34:52 on 2006 September 21.
The follow-up observations performed by {\it Swift} XRT found a remarkable result
in which the quiescent emission of post-burst became 190 times brighter than that of
pre-burst \citep{campana2006}.
In addition to these objects, 
AX\,J1818.8$-$1559 discovered by ASCA \citep{sugizaki2001}
recently 
exhibited a short burst \citep{mereghetti2007b} similar
to those from the magnetar candidates.
Therefore AX\,J1818.8$-$1559 could be a new SGR or AXP \citep{mereghetti2007b}.

The most exciting phenomena among the magnetar candidates would be 
a sudden release of huge energy in rather short period,
the so-called {\it giant flares} from the SGRs.
They typically have a short intense spike
which last less than 1\,s, and
followed by a long pulsating tail
which lasts a few hundred seconds.
Their peak energy flux
can be larger than $\sim10^{6}$ times
Eddington luminosity.
Theoretical studies suggested that the giant flares
were triggered by a catastrophic deformation of
the neutron star crust
due to a torsion of
the strong magnetic field (e.g., \cite{thompson2001}).
Some different emission mechanisms
have been proposed by several authors
\citep{yamazaki2005, lyutikov2006, cea2006}.
In the past three decades, three giant flares were recorded.
The first detection, from the source now known as SGR\,0526$-$66 in
Large Magellanic Cloud (LMC),
was made on March 5 in 1979
\citep{mazets1979, cline1980, evans1980, fenimore1996}.
The second one from SGR\,1900$+$14 was recorded on August 27 in 1998
\citep{hurley1999b, feroci1999, mazets1999, feroci2001, tanaka2007}.
More recently, the most energetic giant flare from SGR\,1806$-$20
was observed on December 27 in 2004
\citep{cameron2005, gaensler2005, hurley2005, mazets2005, palmer2005, terasawa2005, tanaka2007}.
The fluence of its initial intense spike with 600\,ms was 
evaluated to be
$\sim 2$\,erg\,cm$^{-2}$ by the plasma particle detectors on the
Geotail space probe \citep{terasawa2005}.

Soft X-ray spectra of the quiescent emission of the SGRs and
the AXPs were observed by a number of satellites.
Although their spectral model is still under discussion, 
two two-component models are proposed.
One of them is a photoelectrically absorbed two blackbody function (2BB).
Spectral parameters of 2BB are reported by some authors
for the SGRs \citep{mereghetti2006a}
and the AXPs \citep{tiengo2002, morii2003, gotthelf2004, gotthelf2005, halpern2005, tiengo2005, israel2006, gotthelf2007}.
Typical lower and higher temperatures are $\sim 0.5$\,keV and $\sim 1.4$\,keV,
respectively.
The other model is
a photoelectrically absorbed power law plus a blackbody (PL$+$BB).
Some authors report spectral parameters of PL$+$BB
for the SGRs \citep{marsden2001, kurkarni2003, mereghetti2005, mereghetti2006a, mereghetti2007a}
and the AXPs \citep{morii2003, patel2003, rea2003, gotthelf2004, mereghetti2004, woods2004, tiengo2005, gavriil2006, israel2006}.
A typical power law index and a blackbody temperature
are $\sim 3$ and $\sim 0.5$\,keV.
At present it is still unclear which model is more reliable
or physically suitable.

Recent observations by
the International Gamma-Ray Astrophysics Laboratory (INTEGRAL)
discovered a non-thermal hard component in
the spectra of the quiescent emission
above 20\,keV for 5 magnetar candidates \citep{molkov2005, gotz2006b, kuiper2006}.
The non-thermal hard component is well reproduced by a power law model,
$E^{-\Gamma}$, 
where $\Gamma$ ranges
from 1.0-1.8, while the soft X-ray emission below
$\sim12$\,keV, mentioned above, clearly indicates steeper power-law index 
of $\sim 3$ if
the PL$+$BB is applied as the model spectrum.  
Hence, the non-thermal hard emission seen by INTEGRAL is a different 
component and presumably
has a different origin than the soft X-ray emission.
Since some magnetar candidates have two different emission
mechanisms, there seems to be more complex physics than expected before.
Moreover, the non-thermal hard component shows pulsations for 
three AXPs, 1RXS\,J170849.0$-$400910, 4U\,0142$+$614 and
1E\,1841$-$045, through the INTEGRAL and RXTE observations \citep{kuiper2006},
which is related to a neutron star rotation, and hence 
there are particle acceleration processes 
in the vicinity of neutron stars \citep{kuiper2006}.

If the energy source of 
the quiescent emission and the
 bursts 
is the magnetic field as thought to be, 
at least very similar physical process would govern both of them
and their spectra could emerge alike.
It is claimed based on High Energy Transient Explorer 2 (HETE-2) data
that the most acceptable spectral model of the short 
bursts from two SGRs 1806¡Ý20 and 1900+14 is 2BB 
even though it should be regarded just as an empirical model  (Nakagawa 
et al. 2007). 
It would be also preferred to represent spectra of quiescent emissions 
by 2BB rather than BB+PL for SGRs, and even for AXPs if it is  the same class of object.
In this paper, we present a comprehensive spectral study with 2BB
for both the quiescent emission and the for the magnetar candidates.

\section{Data Analyses of Magnetar Candidates}
\subsection{Samples of Magnetar Candidates for Spectral Analyses}
Among magnetar candidates, three SGRs (0526$-$66, 1627$-$41 and 1806$-$20),
one SGR candidate (GRB/SGR\,050925),
six AXPs (CXOU\,J010043.1$-$721134, 4U\,0142$+$614, CXOU\,J164710.2$-$455216,
1RXS\,J170849.0$-$400910, 1E\,1841$-$045 and 1E\,2259$+$586)
and a possible SGR or AXP candidate AX\,J1818.8$-$1559
were used in our study.
For the sake of convenience, we shall refer to
GRB/SGR\,050925 and AX\,J1818.8$-$1559
according to the naming convention for SGRs,
i.e., SGR\,2013$+$34 and SGR\,1819$-$16.
Since results of spectral analyses with
a photoelectrically absorbed two blackbody function (2BB)
using data derived from XMM-Newton observations are reported for SGR\,1900$+$14
\citep{mereghetti2006a}, AXP\,1E\,1048.1$-$5937 \citep{tiengo2005} and
AXP\,XTE\,J1810$-$197 \citep{gotthelf2004, gotthelf2005},
these data were not analyzed in our study.
Two SGR candidates (1801$-$23 and 1808$-$20)
were not also included in the analysis
nor the X-ray counterpart of AXP\,AX\,J1845$-$0258
\citep{tam2006} since its location is still uncertain.
An AXP candidate CXOU\,J160103.1$-$513353
was also not utilized,
because {\it Chandra} observations are not archived at this point.
Table \ref{tab:magnetar_list} shows a summary of utilized magnetar candidates
in our study.
In this paper, we analyzed both the quiescent emission and the short bursts
of the magnetar candidates.

Observations of the quiescent emission utilized in our study were
from the European Photon Imaging Camera (EPIC; \cite{turner2001, struder2001})
on-board XMM-Newton \citep{jansen2001},
the Advanced CCD Imaging Spectrometer (ACIS) on-board
{\it Chandra} and the X-ray Telescope (XRT; \cite{burrows2005})
on-board {\it Swift} \citep{gehrels2004}.
The short bursts were observed by the Burst Alert Telescope (BAT, \cite{barthelmy2005}).
Among the three observational modes (pointing, slew and settling phase) of
{\it Swift} XRT
\citep{capalbi2005} \footnote{This document is available at http://heasarc.gsfc.nasa.gov/docs/swift/analysis/xrt\_swguide\_v1\_2.pdf.},
only data in pointing mode were utilized.
The data observed by XMM-Newton and {\it Chandra} with a timing mode,
and by {\it Chandra} with a grating mode
were not used to reduce spectral uncertainties.

\subsection{Distances to Magnetar Candidates}
Despite careful measurements by many satellites and ground telescopes,
distances to the SGRs and the AXPs are still very uncertain.
In this paper, we used the distances in table \ref{tab:magnetar_list}
(see \cite{woods2006} and references therein).
\citet{park2006} suggested $d \sim 5$\,kpc
for AXP\,CXOU\,J160103.1$-$513353, if this source is associated with
a SNR G330.2$+$1.0.
If AXP\,CXOU\,J164710.2$-$455216 is related to the cluster Westerlund\,1,
the distance might be $d \sim 5$\,kpc \citep{muno2006}.
Since there are no measurements of the distances to
two SGRs 2013$+$34 and 1819$-$16, their distances are
assumed to be $d = 10$\,kpc.
The blackbody radii in this paper were calculated
using the distances in table \ref{tab:magnetar_list}.

\subsection{Data Reductions}\label{data_reduction}
\subsubsection{XMM-Newton}\label{xmm}
The data reductions for XMM-Newton observations
were made using the SAS 7.0.0 software in the following way. 
To apply the latest calibration results to the data,
the basic pipeline processing
using the SAS tasks {\it emchain} and {\it epchain}
were performed.
Proton flares are usually seen in
the light curves of these observations
\citep{snowden2004}\footnote{This document is available at ftp://legacy.gsfc.nasa.gov/xmm/doc/xmm\_abc\_guide.pdf}.
If the count rate of the proton flare is large
compared with the nominal background,
the proton flare cannot be
ignored in the spectral analyses.
Therefore those high background time regions
were excluded using a threshold of twice the nominal background.
The effects of photon pile-up
were investigated using {\it epatplot}
for each observation.
The circular foreground and background regions
were determined by eye and their spectra
were extracted using {\it xmmselect}.
The response matrix files and the auxiliary response files
were calculated
using {\it rmfgen} and {\it arfgen}.
We only considered phptons in the 0.6-12\,keV band.

\subsubsection{Chandra}\label{cxo}
The data reductions for {\it Chandra} observations
were made using the CIAO 3.3 software in the following way.
Since some observations were not applied
to the latest calibration results,
the new Level=2 event files
were created for these observations.
First, the {\it acis\_detect\_afterglow} corrections
were removed using the CIAO tool {\it dmtcalc}.
After that, hot pixels and cosmic ray afterglows
were identified using {\it acis\_run\_hotpix}.
Then, the new Level=2 event files
were created using {\it acis\_process\_events}.
Using the data applied
to the latest calibration results,
the circular foreground and background regions
were determined by eye, and their spectra
were extracted using {\it dmextract}.
The response matrix files
were calculated using {\it acis\_fef\_lookup} and {\it mkrmf}.
The auxiliary response files
were calculated using {\it asphist} and
{\it mkarf}.
We only considered phptons in the 0.1-6\,keV band.

\subsubsection{Swift BAT}
The data reductions for {\it Swift} observations
were made using the HEAsoft 6.1.1 software.
For the data reductions of {\it Swift} BAT,
the following steps were performed.
Light curves in 15-150 keV
were generated using {\it batbinevt},
and the foreground time regions
were determined by eye for each burst.
The foreground spectra
were generated with those time regions
using {\it batbinevt}.
The background was already subtracted using a mask weighting
method by the {\it Swift} Data Center.
Corrections due to the spacecraft slewing during a burst
were
applied to the foreground spectral files using {\it batupdatephakw}.
The systematic errors recommended by the
BAT team were applied to the
foreground spectral files using {\it batphasyserr}.
The response matrix files
were calculated using {\it batdrmgen}.
We only used photons in the 15-15\,keV band.

\subsubsection{Swift XRT}\label{swift_xrt}
To reduce the {\it Swift} XRT data we
performed the following.
We generated the Level 2 (screened) event files
using the pipeline processing tool {\it xrtpipeline}.
The effects of photon pile-up
were investigated following the methods described
by \citet{romano2006}.
The foreground and background regions
were determined with a rectangle
for WT mode and a circle for PC mode
selected by eye.
The spectra were
extracted using XSELECT V2.3 which is
part of the HEAsoft 6.1.1 software
package.
The auxiliary response files
were calculated using {\it xrtmkarf}.
The response matrix files included in CALDB 20060407
were utilized.
We only considered photons in the 0.6-10\,keV band.

\subsection{Observations and Spectral Analyses}\label{sgr_axp_result}
A recent study using HETE-2 data by \citet{nakagawa2007}
revealed that the most acceptable spectral model for the
SGR short bursts is a 2BB.
If the
 bursts and the quiescent emission are both
activated by magnetic dissipation,
the quiescent emission spectra may be reproduced by the same
spectral model for the
 bursts (i.e., 2BB).
Then we performed the spectral analyses with
a 2BB model
for both the short bursts and the quiescent emission for all samples.
If reliable temperatures and/or radii were
not determined using 2BB
because of insufficient statistics due to small exposure times
and/or a faint object,
a photoelectrically absorbed single blackbody (BB) was used.
Note that because
the spectral analyses of the short bursts detected by {\it Swift} BAT
were performed using data above 15\,keV,
a photoelectric absorption was not required.

Photon pile-up effect
was negligible for all observations utilized here.
We used shapes and sizes of the foreground and background regions
presented in subsection \ref{data_reduction}
and tables \ref{tab:sgr_obs_xmm}-\ref{tab:axp_j1647_obs_swift},
respectively.
The spectral fits
were
performed with XSPEC 12.3.0 \citep{arnaud1996}
in HEAsoft 6.1.1 software.
The spectra were
binned to at least 25 counts in each spectral bin
using {\it grpppha}.
The spectral parameters are summarized in table \ref{tab:sgr_spc_xmm}-\ref{tab:axp_j1647_spc_swift},
where the quoted errors are 68 \% confidence level for fluxes and 90 \% confidence
level for other parameters.

\subsubsection{SGR\,0526$-$66}
SGR\,0526$-$66 was observed at three
epochs by {\it Chandra} from 2000 to 2001.
Two observations were utilized in our study (table \ref{tab:sgr_obs_cxo}),
because one observation on 2001 September 1 was affected by a photon pile-up
effect \citep{kurkarni2003}.
The spectra were well reproduced by 2BB (table \ref{tab:sgr_spc_cxo}).

Two XMM-Newton observations from 2000 to 2001 were not utilized,
because we could not distinguish between SGR\,0526$-$66 and its SNR.

\subsubsection{SGR\,1627$-$41}
SGR\,1627$-$41 was observed at three
epochs by XMM-Newton in 2004 (table \ref{tab:sgr_obs_xmm}).
Since the spectral fits with 2BB could not give
reliable lower and higher temperatures
($kT_{\mathrm{LT}}$ and $kT_{\mathrm{HT}}$),
BB was used (table \ref{tab:sgr_spc_xmm}).

This object was also observed at 4 epochs by {\it Chandra} from 2001 to 2005.
Two observations were utilized (table \ref{tab:sgr_obs_cxo}),
because one observation on 2003 April 21 was performed with a timing mode
and one observation on 2005 June 28 had less statistics due to
small net exposure time (9.83\,ks).
The spectra were reproduced by BB (table \ref{tab:sgr_spc_cxo}),
because reliable $kT_{\mathrm{LT}}$ and $kT_{\mathrm{HT}}$ could not be obtained
by 2BB.

\subsubsection{SGR\,1806$-$20}
SGR\,1806$-$20 was observed at
6 epochs by XMM-Newton from 2003 to 2005 (table \ref{tab:sgr_obs_xmm}).
Among them, 4 observations
were performed before the giant flare on 2004 December 27,
while the other two observations were performed after it.
The spectra were well reproduced by 2BB (table \ref{tab:sgr_spc_xmm}).

This object was also observed at 11 epochs by {\it Chandra}
from 2000 to 2006.
The {\it Chandra} observations were not utilized in our study,
because three imaging observations were affected by a photon pile-up effect
\citep{kaplan2002}.

\subsubsection{Burst on 2005 September 25 - A Candidate of SGR}
We analyzed the short burst from SGR\,2013$+$34 \citep{holland2005}.
The burst spectrum was generated
using the {\it Swift} BAT data from 0.06\,s to 0.17\,s
and well reproduced by 2BB (table \ref{tab:sgr_050925_spc_swift}).

The X-ray counterpart was observed at 4 epochs by {\it Swift} XRT from 2005 to 2006.
These observations were not utilized in our study,
because we could not find the X-ray counterpart or there was practically
no exposure time for the WT mode.
For PC mode, there was not
enough statistics to perform the spectral analyses.

The follow-up observation by XMM-Newton (table \ref{tab:sgr_obs_xmm})
also detected the X-ray counterpart \citep{luca2005}.
The spectrum was reproduced by BB (table \ref{tab:sgr_spc_xmm}),
because reliable spectral parameters were not obtained by 2BB.

\subsection{Burst on 2007 October 17 - A Candidate of SGR or AXP}
We analyzed the short burst from SGR\,1819$-$16 \citep{mereghetti2007b}.
This object was observed at one epoch by XMM-Newton in 2003 (table \ref{tab:sgr_obs_xmm}).
Since reliable spectral parameters were not obtained by 2BB,
BB was used (table \ref{tab:sgr_spc_xmm}).
In table \ref{tab:sgr_spc_xmm},
the temperature is a little bit larger than
a typical value for the SGRs and AXPs (see figure \ref{fig:bb_relation})
as already reported by \citet{tiengo2007}.
Note that a photoelectrically absorbed power law model (PL)
also gave an acceptable result of $\chi^2/\mathrm{d.o.f.} = 56/74$,
which is consistent with a result reported by \citet{tiengo2007}.
Further observations should be encouraged to reveal
whether SGR\,1819$-$16 is a new SGR or AXP.

\subsubsection{AXP\,CXOU\,J010043.1$-$721134}
AXP\,CXOU\,J010043.1$-$721134 was observed at three
epochs by XMM-Newton from 2000 to 2005.
One observation on 2005 March 27 was not utilized in our study,
because the pn camera was not operated,
and the object fell on a gap of the CCD chips for the MOS1 and MOS2 cameras.
For the other two observations (table \ref{tab:axp_obs_xmm}),
the data of the pn and MOS1 cameras were utilized in our study,
because the object fell on a gap of the CCD chips for the MOS2 camera.
The spectra were well reproduced by 2BB (table \ref{tab:axp_spc_xmm}).

AXP\,CXOU\,J010043.1$-$721134 was also observed at
6 epochs by {\it Chandra}
from 2001 to 2004 (table \ref{tab:axp_obs_xmm}).
The spectra for three observations were well reproduced by 2BB,
while the spectra of the other three observations were reproduced by BB
because reliable $kT_{\mathrm{LT}}$ could not be obtained by 2BB (see table \ref{tab:axp_spc_cxo}).

\subsubsection{AXP\,4U\,0142$+$614}\label{obs_spc_4u_0142}
AXP\,4U\,0142$+$614 was observed at
4 epochs by XMM-Newton from 2002 to 2004.
One observation on 2002 February 13 was not utilized in our study,
because the background level
became 10 times higher than the ordinary background level \citep{gohler2005}.
We just utilized one observation on 2003 January 24
by the pn camera (table \ref{tab:axp_obs_xmm}),
because the other observations were performed with a timing mode
or affected by pile-up.
The spectrum was not reproduced by a
2BB ($\chi^2/\mathrm{d.o.f.} = 1086/819$), because
there seems to be an excess above $\sim$\,7\,keV (table \ref{tab:axp_spc_xmm}).

This object was also observed at 4 epochs by {\it Chandra} from 2000 to 2006.
Although three observations were archived, we did not utilize them
because they were affected by pile-up.

\subsubsection{AXP\,CXOU\,J164710.2$-$455216}
We analyzed the short burst from AXP\,CXOU\,J164710.2$-$455216 \citep{krimm2006}.
We generated the burst spectrum
using the {\it Swift} BAT data from
$t = 0.0585$\,s to $t = 0.0725$\,s, where $t = 0 $ indicates the trigger time.
Since reliable $kT_{\mathrm{LT}}$ and $R_{\mathrm{LT}}$ could not be obtained by 2BB,
BB was used (table \ref{tab:axp_j1647_spc_swift}).

The post-burst emission was observed 15 times by {\it Swift} XRT
from 2006 to 2007.
We performed the joint spectral analyses using the data in both WT and PC modes
for four observations,
and their spectra were well reproduced by 2BB (table \ref{tab:axp_j1647_spc_swift}).
On the other hand, the spectra of the other 12 observations were
fitted by BB (table \ref{tab:axp_j1647_spc_swift}),
because reliable $kT_{\mathrm{LT}}$ and $R_{\mathrm{LT}}$
could not be achieved by 2BB.

The quiescent emission was observed at 7 epochs by {\it Chandra} from 2005 to 2007.
The observations were performed before the short burst,
while the other 5 observations were performed after it.
The two pre-burst observations (table \ref{tab:axp_obs_cxo}) were utilized,
because the post-burst observations were performed
in timing mode.
The spectral analyses by BB gave
a rather large reduced $\chi^{2}$
which were consistent with \citet{muno2006}.
The spectral analyses were improved using 2BB,
but a reliable $kT_{\mathrm{HT}}$ could not be obtained.
In table \ref{tab:axp_spc_cxo}, we report the results of the spectral analyses by BB.

Although the quiescent emission of the post-burst was observed at one
epoch by {\it Suzaku} in 2006
\citep{naik2008},
we did not utilize this data in our study.
The two follow-up observations by XMM-Newton were not utilized,
because they were not archived at this point.

\begin{figure}
  \begin{center}
    \FigureFile(80mm,129mm){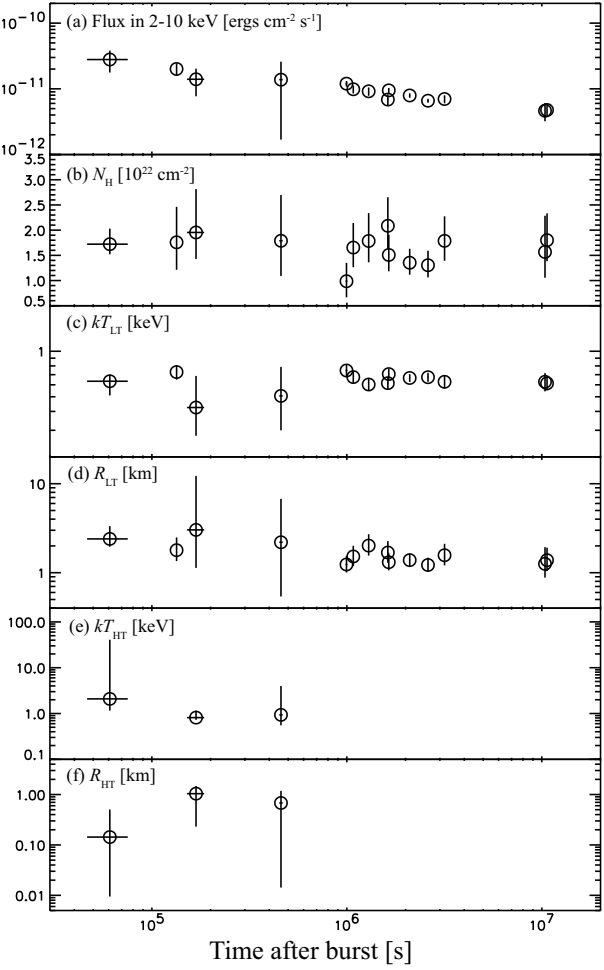}
  \end{center}
  \caption{Time history of the (a) flux in 2-10 keV in units of ergs cm$^{-2}$ s$^{-1}$,
  (b) photoelectric absorption $N_{\mathrm{H}}$ in units of cm$^{-2}$,
  (c) temperature of the lower blackbody $kT_{\mathrm{LT}}$ in units of keV,
  (d) radius of the lower blackbody $R_{\mathrm{LT}}$ in units of km,
  (e) temperature of the higher blackbody $kT_{\mathrm{HT}}$ in units of keV
  and (f) radius of the higher blackbody $R_{\mathrm{HT}}$ in units of km
  for the emissions of AXP\,CXOU\,J164710.2$-$455216
  observed by XRT/{\it Swift}.}\label{fig:lc_persistent_cxou_j16}
\end{figure}

Figure \ref{fig:lc_persistent_cxou_j16} shows a time history of the spectral parameters
obtained by the {\it Swift} XRT data and the flux is decaying in time.

\subsubsection{AXP\,1RXS\,J170849.0$-$400910}\label{obs_spc_1rxs_j17}
AXP\,1RXS\,J170849.0$-$400910 was observed at one
epoch by XMM-Newton in 2003 (table \ref{tab:axp_obs_xmm}).
Since pile-up for the MOS1 and MOS2 cameras
was significant, only the pn camera data was utilized.
The spectrum was not reproduced by a
2BB ($\chi^2/\mathrm{d.o.f.} = 1566/1232$),
there seemed to be an excess above $\sim$\,7\,keV (table \ref{tab:axp_spc_xmm}),
similar to
AXP\,4U\,0142$+$614 (see subsubsection \ref{obs_spc_4u_0142}).

\subsubsection{AXP\,1E\,1841$-$045}
AXP\,1E\,1841$-$045 was observed at two
epochs by XMM-Newton in 2002 (table \ref{tab:axp_obs_xmm}).
The spectra were well reproduced by 2BB (table \ref{tab:axp_spc_xmm}).

This object was also observed by {\it Chandra} in 2000
and the detailed spectral analyses with 2BB were reported by \citet{morii2003}.
Therefore we did not utilize this observation.

\subsubsection{AXP\,1E\,2259$+$586}\label{obs_spc_1e_2259}
AXP\,1E\,2259$+$586 was observed at 6 epochs by XMM-Newton
from 2002 to 2005.
We just utilized three observations by the pn camera (table \ref{tab:axp_obs_xmm}),
because the other observations were affected by
a photon pile-up effect.
The spectra were not reproduced by 2BB for two observations
($\chi^2/\mathrm{d.o.f.} = 1167/888$ and $\chi^2/\mathrm{d.o.f.} = 1531/1053$),
there seemed to be an excess above $\sim$\,7\,keV (table \ref{tab:axp_spc_xmm}).
These results are the same as those of
AXPs 4U\,0142$+$614 and 1RXS\,J170849.0$-$400910
(see subsubsections \ref{obs_spc_4u_0142} and \ref{obs_spc_1rxs_j17}).
Note that the spectrum of one observation on 2005 July 28 was well reproduced
by 2BB ($\chi^2/\mathrm{d.o.f.} = 457/458$ in table \ref{tab:axp_spc_xmm})
in spite of a small net exposure time (2.66\,ks).

This object was also observed at two epochs by {\it Chandra} in 2000 and 2006.
These observations were not utilized in our study,
because one observation on 2000 January 12 was affected by
pile-up \citep{patel2001} and the other observation
on 2006 May 9 was not archived at this point.

\section{Discussions}
\subsection{Two Possible Spectral Models for Quiescent Emission}
Quiescent emission spectra of the AXPs were used to
examine the blackbody plus power law model (BB$+$PL).
Recent studies suggested that the quiescent emission spectra of the SGRs and AXPs
are reproduced by either a two blackbody function (2BB)
or a BB$+$PL.
To compare 2BB and BB$+$PL, spectral fits for one SGR\,1806$-$20 observation (0205350101)
and one AXP\,4U\,0142$+$614 observation (0112781101) were performed using these two
spectral models.
For SGR\,1806$-$20, a spectral fit with 2BB gives $\chi^{2}$/d.o.f. = 2244/2224 with $P = 0.38$,
while a spectral fit with BB$+$PL gives $\chi^{2}$/d.o.f. = 2274/2224 with $P = 0.23$.
Here, $P$ denotes a null hypothesis probability.
For AXP\,4U\,0142$+$614, a spectral fit with 2BB gives $\chi^{2}$/d.o.f. = 1086/819 with $P = 10^{-10}$,
while a spectral fit with BB$+$PL gives $\chi^{2}$/d.o.f. = 978/819 with $P = 10^{-5}$.
In this case, both spectral models are rejected.
These three datasets have all good statistics, 
therefore these acceptability and unacceptability of the fits
are not simply due to their statistics, and may reflect 
complexity of spectral shape of radiations from these SGRs and AXPs.
A recent study using HETE-2 data reports that the most acceptable model of
SGR short burst spectra is 2BB even if it is an empirical model \citep{nakagawa2007}.
It is very interesting to investigate if
the spectra of both the quiescent emission and the
bursts are reproduced by same spectral model 2BB.

\subsection{Spectral Parameter of Two Blackbody Function}
As shown in subsection \ref{sgr_axp_result}, 
both the spectra of the quiescent emission
and short bursts
were well reproduced by 2BB with some exceptions.
The quiescent emission spectra of three AXPs
(4U\,0142$+$614, 1RXS\,J170849.0$-$400910 and
1E\,2259$+$586) seemed
to have an excess above $\sim7$\,keV (see subsection \ref{excess}).
In some cases, the spectra of the quiescent emission
and short bursts
were fitted with a photoelectrically absorbed single blackbody function (BB).
This was just due to a low
X-ray flux and/or insufficient exposure time
to determine the reliable 2BB spectral parameters.
Therefore we restrict our discuss to spectra
that were well modeled by a 2BB model in order
to investigate
the global characteristic among the magnetar candidates, and between 
the quiescent emission and the bursts.

\begin{figure}
  \begin{center}
    \FigureFile(80mm,83mm){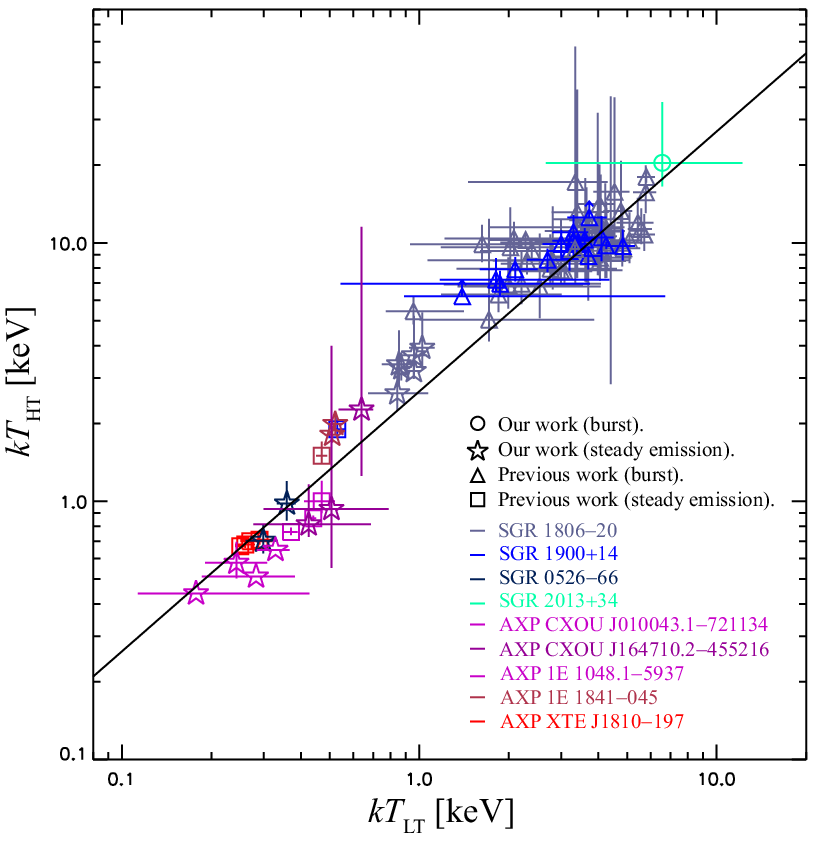}
  \end{center}
  \caption{Relationship between the 2BB temperatures $kT_{\mathrm{LT}}$ and $kT_{\mathrm{HT}}$.
  The triangles and squares denote the previous work on
  the bursts \citep{feroci2004, olive2004, gotz2006a, nakagawa2007} and
  the quiescent emission
  \citep{morii2003, gotthelf2004, gotthelf2005, tiengo2005, mereghetti2006a}, respectively.
  The circles and stars denote our work on the bursts
  and the quiescent emission, respectively.
  The line represents the best-fit power law model.}\label{fig:bb_relation}

\end{figure}

Figure \ref{fig:bb_relation} shows the relationship between lower
and higher temperatures ($kT_{\mathrm{LT}}$ and $kT_{\mathrm{HT}}$).
The 2BB temperatures were
obtained by our study and previous works
\citep{morii2003, feroci2004, olive2004, gotthelf2004, gotthelf2005,
tiengo2005, gotz2006a, mereghetti2006a, nakagawa2007}.
The 2BB temperatures of 51 short bursts detected by HETE-2 are also
plotted in figure \ref{fig:bb_relation} \citep{nakagawa2007}.

In figure \ref{fig:bb_relation}, there seems to be a strong correlation
between $kT_{\mathrm{LT}}$ and $kT_{\mathrm{HT}}$.
It is remarkable that the correlation seems to be
independent of the objects and/or the emission types
(the quiescent emission or the burst).
In order to clarify the correlation, it is essential to
consider systematic errors among
the satellites.
The systematic errors of fluxes between XMM-Newton and {\it Chandra} are
reported to be 10-20\% \citep{snowden2002}\footnote{The other document is available at http://cxc.harvard.edu/ccw/proceedings/04\_proc/presentations/kashyap/kashyap.pdf.}.
Since the data of these satellites are important for the correlation,
we focus our attention on the above systematic errors of 15\%.
The amount errors are described by $\sqrt{\sigma_{\mathrm{1}}^2 + \sigma_{\mathrm{2}}^2}$,
where $\sigma_{\mathrm{1}}$ and $\sigma_{\mathrm{2}}$ are
statistical and systematic errors, respectively.
The $kT_{\mathrm{LT}}$-$kT_{\mathrm{HT}}$ relation
was fitted with a power law model
$kT_{\mathrm{HT}} = A\left({kT_{\mathrm{LT}}}\right)^{\eta}$,
where $A$ is a normalization and $\eta$ is an index.
Here, the errors on $kT_{\mathrm{HT}}$ were taken into account
for the fitting.
In the following discussions, quoted errors on all parameters are
90\% confidence level.
The parameters were found to be
$A = 2.7\pm1.1$ and $\eta = 1.0\pm0.3$
with $\chi^2/\mathrm{d.o.f.} = 86/92$.
Interestingly, the derived index is just unity;
this implies that
$kT_{\mathrm{LT}}$ and $kT_{\mathrm{HT}}$ have a {\it linear} correlation
and the ratio
$kT_{\mathrm{LT}}/kT_{\mathrm{HT}} = 0.37\pm0.15$ is almost
constant over 1.5 order of magnitudes.
The linear correlation coefficient between $kT_{\mathrm{LT}}$
and $kT_{\mathrm{HT}}$ was $r = 0.99$.
In addition, the relation was separately determined 
for the quiescent emission and the burst,
where the parameters were
$A = 2.6\pm0.9$ and $\eta = 1.0\pm0.3$ with $\chi^2/\mathrm{d.o.f.} = 28/27$,
and $A = 5.8\pm2.5$ and $\eta = 0.4_{-0.3}^{+0.5}$ with 
$\chi^2/\mathrm{d.o.f.} = 18/63$,
respectively.
The linear correlation coefficient was $r = 0.97$ for both cases.
The index for the quiescent emission is consistent with
that for the burst within 90\% confidence level.
This means that $kT_{\mathrm{LT}}$ and $kT_{\mathrm{HT}}$ are
very well correlated
irrespective of objects and/or emission types (i.e., the quiescent emission
and the bursts).
Note that the fitting with the errors on $kT_{\mathrm{LT}}$ gave
consistent results in each model fitting.
It is also worth noting that the spectrum of a tentative magnetar candidate,
1E\,1207$-$5209,
was well reproduced by 2BB \citep{luca2004}, and the ratio
$kT_{\mathrm{LT}}/kT_{\mathrm{HT}} = 0.514\pm0.004$
is marginally consistent with the ratio of the magnetar candidates.

\begin{figure}
  \begin{center}
    \FigureFile(80mm,80mm){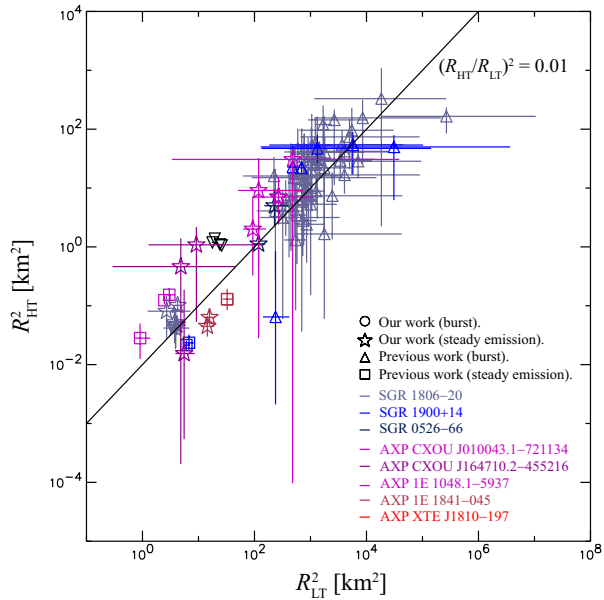}
  \end{center}
  \caption{Relationship between the square of the blackbody radii
  $R^2_{\mathrm{LT}}$ and $R^2_{\mathrm{HT}}$.
  The triangles and squares denote the previous work on
  the the bursts \citep{olive2004, nakagawa2007} and
  the quiescent emission
  \citep{morii2003, tiengo2005, mereghetti2006a}, respectively.
  The stars denote our work on the quiescent emission.
  The solid line shows a ratio of $R_{\mathrm{HT}}^{2}$ to $R_{\mathrm{LT}}^{2}$
  of 0.01.}\label{fig:radius_relation}
\end{figure}

The constant ratio $kT_{\mathrm{LT}}/kT_{\mathrm{HT}} \approx 0.4$
may imply that the spectra of the magnetar candidates
have a similar shape, the emission radii should be considered.
Indeed, there seems to be a correlation
between $R_{\mathrm{LT}}^2$ and $R_{\mathrm{HT}}^2$ (figure \ref{fig:radius_relation}).
The 2BB radii are derived from our study and the previous works.
The solid line in figure \ref{fig:radius_relation} shows that a ratio
defined as $(R_{\mathrm{LT}}/R_{\mathrm{HT}})^2$ get constant value (0.01).

The linear correlations of $kT_{\mathrm{LT}}$-$kT_{\mathrm{HT}}$ and
$R_{\mathrm{LT}}^2$-$R_{\mathrm{HT}}^2$ might imply that
all the spectra have similar shape.
In other words, there might be the
same emission mechanisms among the magnetar candidates,
and between the quiescent emission and the bursts even
though 2BB is an empirical model.
The latter is reminiscent of the
relationship between frequent microflares
and ordinary solar flares of the
sun \citep{feldman1995, shimizu1995, yuda1997}.
The microflares are dim, small scale flares, while the solar flares are
bright, large scale flares.
The microflares are thought to play an important role to heat
the solar corona.
Recently, it was revealed
by {\it Hinode} \citep{ichimoto2005}
that the microflares occurred in the active
bright regions on the surface of the sun.
Considering the relationship between the microflares and the solar flares, 
the quiescent emissions of the magnetar candidates
could be due to frequent small scale activity.
On the other hand, the could be due to
larger scale activity
(or an avalanche like event of the small scale activity).

In figure \ref{fig:bb_relation}, $kT_{\mathrm{HT}}$ of the quiescent
emission for some magnetar candidates clearly exceeds $\sim$2\,keV.
Using $kT_{\mathrm{HT}}$ and $R_{\mathrm{HT}}$ obtained from the
observation of 0148210101 for SGR\,1806$-$20, a flux of the $kT_{\mathrm{HT}}$
component turned out to be
$F_{\mathrm{HT}} = {4.84\times10^{25}}{(kT_{\mathrm{HT}}/2.62\,\mathrm{keV})^4}$\,ergs\,cm$^{-2}$\,s$^{-1}$,
larger than the Eddington flux
$F_{\mathrm{Edd}} = {1.34\times10^{25}}{(M_{\mathrm{NS}}/1.4\,\MO)(R_{\mathrm{NS}}/10\,\mathrm{km})^{-2}}$\,ergs\,cm$^{-2}$\,s$^{-1}$,
where $M_{\mathrm{NS}}$ is the mass of the
neutron star and $R_{\mathrm{NS}}$ is
the radius of the neutron star.
This implies that a radiation pressure is very strong and the plasma of
the $kT_{\mathrm{HT}}$ component cannot exist steadily.
This could be due to the combined effects of the confinement of the
plasma and the strong magnetic field surpressing
the motion
of the particles in directions perpendicular to the field lines,
thus decreasing the cross-sectin for Compton scattering.

One can see large emission radii of $R_{\mathrm{LT}}^2 \gtrsim 100^2$\,km
for the bursts in figure \ref{fig:radius_relation}.
The magnetic field, $B$, is dramatically decreased
as one moves from the center of
the neutron star, $R$,
because $B \propto R^{-3}$.
The plasma of the $kT_{\mathrm{LT}}$ component might be diffused by radiation pressure
without a magnetic field
as strong as
$\sim10^{15}$\,G at $R \gtrsim 100$\,km.
To investigate whether the plasma is diffused,
the magnetic pressure and the radiation pressure at $R \gtrsim 100$\,km
were investigated.
The magnetic pressure and the radiation pressure at an outer
radius of the emission region were estimated using 
spectral parameters of
the short burst of \#3854
($kT_{\mathrm{LT}} = 1.7$\,keV and $R_{\mathrm{LT}} = 136$\,km)
in \citet{nakagawa2007}.
A spherical emission region was considered for
the sake of simplicity.
The center of the emission region was assumed to be aligned to
the center of the neutron star.
The magnetic pressure turns out to be
$p_{\mathrm{m}}\sim1.6\times10^{21}(R/136$\,km$)^{-6}(B_{\mathrm{s}}/5.0\times10^{14}\,\mathrm{G})^2$\,ergs\,cm$^{-3}$,
where $B_{\mathrm{s}}$ is
the assumed surface dipole magnetic field at $R = 10$\,km.
The radiation pressure turned out to be
$p_{\mathrm{r}}\sim3.8\times10^{14}(kT/1.7$\,keV$)^{4}$\,ergs\,cm$^{-3}$,
where $kT$ is a blackbody temperature of the emission region.
Consequently, the plasma is not diffused by the radiation pressure
because $p_{\mathrm{m}} > p_{\mathrm{r}}$.

\begin{figure}
  \begin{center}
    \FigureFile(160mm,80mm){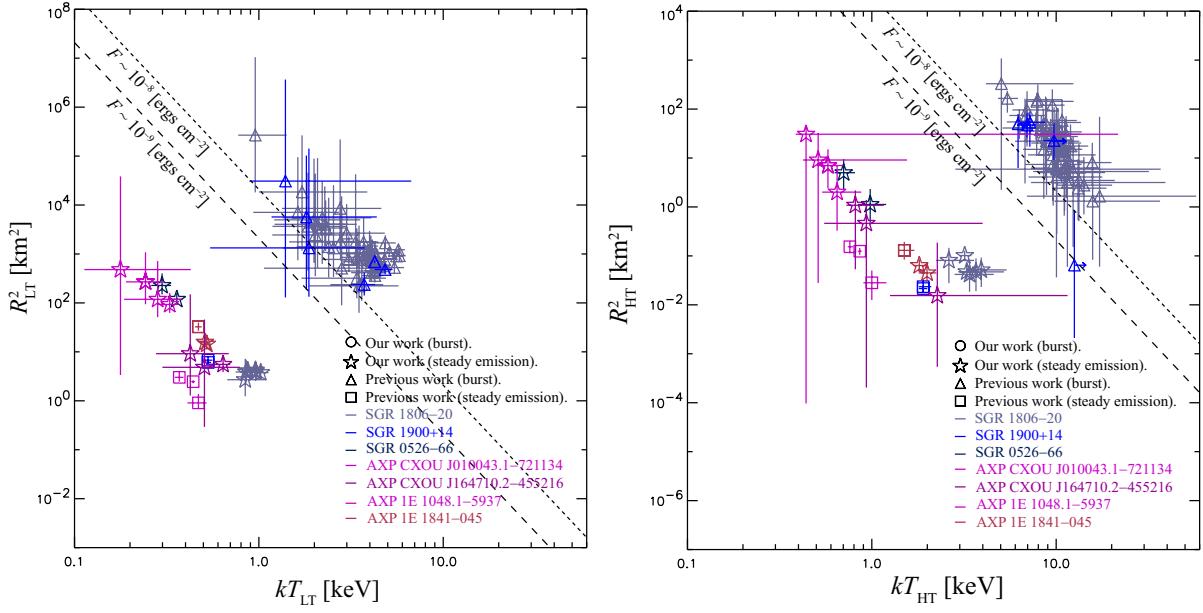}
  \end{center}
  \caption{{\it Left:} Relationship between the lower temperature
  of 2BB $kT_{\mathrm{LT}}$ and the square of the blackbody radii of 2BB $R_{\mathrm{LT}}^2$.
  {\it Right:} Relationship between the higher temperature
  of 2BB $kT_{\mathrm{HT}}$ and the
  square of the blackbody radii $R_{\mathrm{HT}}^2$.
  The dotted and dashed lines correspond to bolometric fluences of $10^{-8}$
  and $10^{-9}$\,ergs\,cm$^{-2}$, respectively.}\label{fig:kt_rsquared_relation}
\end{figure}

Figure \ref{fig:kt_rsquared_relation} shows the relationships
between $kT_{\mathrm{LT}}$ and $R_{\mathrm{LT}}^2$ ({\it left}),
and between $kT_{\mathrm{HT}}$ and $R_{\mathrm{HT}}^2$ ({\it right}).
One may see that data points of the quiescent emission and the burst
are apparently clustering in separate areas of this plot
despite the linear correlations for
$kT_{\mathrm{LT}}$-$kT_{\mathrm{HT}}$ and $R_{\mathrm{LT}}^2$-$R_{\mathrm{HT}}^2$.
The data points seems to be distributed along the direction of two lines
with same functional form
in figure \ref{fig:kt_rsquared_relation}.
This might imply a different origin and/or mechanism for
the quiescent emission and the burst.
Combining the $kT_{\mathrm{LT}}$-$kT_{\mathrm{HT}}$ and $R_{\mathrm{LT}}^2$-$R_{\mathrm{HT}}^2$
correlations, and the above mentioned speculations,
the bolometric luminosity might be given by a function of
$kT_{\mathrm{LT}}$, $kT_{\mathrm{HT}}$, $R_{\mathrm{LT}}$ or $R_{\mathrm{HT}}$
with the same indices irrespective of the emission types (i.e., the quiescent emission and the bursts).
To clarify this hypothesis, relations between the bolometric luminosity
and $kT_{\mathrm{LT}}$, $kT_{\mathrm{HT}}$, $R_{\mathrm{LT}}$ or $R_{\mathrm{HT}}$
were investigated.
We found that the index of the quiescent emission was not consistent with the
index of the burst in any cases.
Therefore, the apparent clustering in separate areas for the
quiescent emission and the burst in figure \ref{fig:kt_rsquared_relation} might not be real.
It is obvious to consider the detectability for the burst with
the instrument of wide field of view, such as the WXM on-board HETE-2,
comparing to that of the narrow field of view detectors with X-ray telescopes
for the quiescent emission.
In figure \ref{fig:kt_rsquared_relation}, the dotted and dashed
lines correspond to bolometric fluences of $10^{-8}$ and $10^{-9}$\,ergs\,cm$^{-1}$, respectively.
Most of the short bursts localized by the WXM/HETE-2 have
fluences greater than $10^{-8}$\,ergs\,cm$^{-2}$ \citep{nakagawa2007}.
This may suggest that dim bursts
($\lesssim10^{-8}$\,ergs\,cm$^{-2}$)
are not detectable by the WXM/HETE-2.
Such dim bursts may fall on a gap between the burst population
and the quiescent emission population.

\subsection{An Excess above $\sim$\,7\,keV for Three AXPs}\label{excess}
Soft X-ray spectra of the quiescent emission of three AXPs 1E\,2259$+$586, 4U\,0142$+$614
and 1RXS\,J170849.0$-$400910 observed by XMM-Newton were not reproduced by 2BB
in spite of enough statistics (see subsection \ref{sgr_axp_result}).
One can see an excess above $\sim$\,7\,keV in their spectra.

Recent studies discovered a non-thermal hard component above $\sim$\,20\,keV 
in the quiescent emission of the SGRs and the AXPs using data derived from INTEGRAL
observations \citep{molkov2005, gotz2006b, kuiper2006}.
The spectra of the non-thermal hard component were well reproduced
by a power law model \citep{molkov2005, gotz2006b, kuiper2006}.
\citet{kuiper2006} reported the non-thermal hard component for
the AXPs 4U\,0142$+$614 and 1RXS\,J170849.0$-$400910,
while they estimated just an upper limit for AXP 1E\,2259$+$586.

\begin{figure}
 \begin{center}
  \FigureFile(80mm,80mm){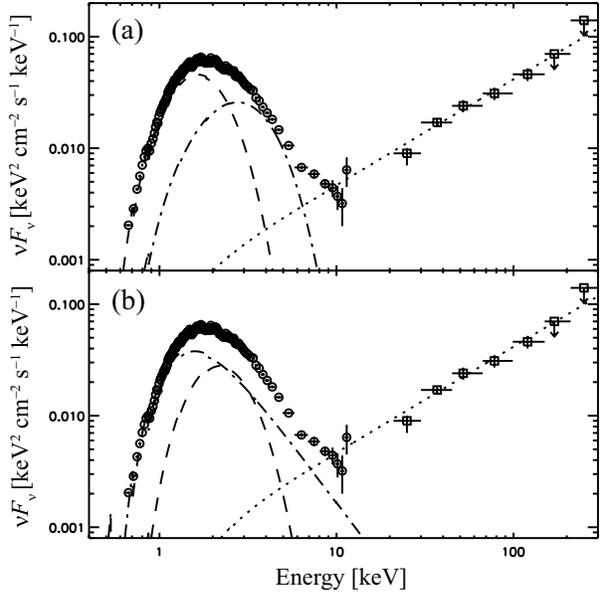}
 \end{center}
 \caption{A schematic view of $\nu{F}_{\nu}$ spectra using 2BB$+$PL (a)
 and BB$+$2PL (b) for AXP\,4U\,0142$+$614. Spectral parameters of
 X-ray spectra (i.e., $\lesssim$12\,keV) are derived from our analyses
 using the XMM-Newton
 observation of 0112781101, while those of the non-thermal hard component
 (i.e., $\gtrsim$20\,keV) is obtained by INTEGRAL observations \citep{kuiper2006}.
 The circles denote data derived from our analyses
 using the XMM-Newton observation of 0112781101.
 The squares represent INTEGRAL observations taken from Fig.7 in
 \citet{kuiper2006} by eye.
 The dashed, dot-dash, dotted lines in (a) show the $kT_{\mathrm{LT}}$ component,
 the $kT_{\mathrm{HT}}$ component and the PL component for the hard spectrum, respectively.
 Those lines in (b) show the $kT$ component, the PL component for an X-ray spectrum
 and the PL component for the hard spectrum, respectively.}\label{fig:eeufspec_summary}
\end{figure}

The excesses above $\sim$\,7\,keV in our data seem to be associated with
the non-thermal hard component above $\sim$\,20\,keV discovered by INTEGRAL.
To investigate our idea,
a power law model related to the non-thermal hard component reported by
\citet{kuiper2006} was
added to the 2BB and BB$+$PL spectral fits.
A photon index and a normalization at 20\,keV of the additional power law model
were fixed to 1.05 and $2.3 \times 10^{-5}$\,photons\,cm$^{-2}$\,s$^{-1}$\,keV
for AXP\,4U\,0142$+$614, 
and 1.44 and $8.8 \times 10^{-6}$\,photons\,cm$^{-2}$\,s$^{-1}$\,keV
for AXP\,1RXS\,J170849.0$-$400910 \citep{kuiper2006}.
For AXP\,4U\,0142$+$614, 2BB$+$PL gave
an acceptable result of
$\chi^2/\mathrm{d.o.f.} = 876/819$ with $P = 0.08$
while BB$+$2PL gave a poor result of
$\chi^2/\mathrm{d.o.f.} = 1067/819$ with $P = 1\times10^{-8}$.
On the other hand, 2BB$+$PL gave a poor result of
$\chi^2/\mathrm{d.o.f.} = 1438/1232$ with $P = 4\times10^{-5}$
while BB$+$2PL gave
$\chi^2/\mathrm{d.o.f.} = 1270/1232$ with $P = 0.22$
for AXP\,1RXS\,J170849.0$-$400910.
In figure \ref{fig:eeufspec_summary} (a), the schematic view of a 2BB$+$PL spectrum
for AXP\,4U\,0142$+$614 is represented.
The circles and squares
denote observational data,
while the dashed, dot-dash and dotted lines are model.
For the sake of comparison,
the schematic view of a BB$+$2PL spectrum for AXP\,4U\,0142$+$614
is also shown in figure \ref{fig:eeufspec_summary} (b).
Note that a distinctive hard component (e.g., the harder power law)
is required for either case to represent non-thermal component seen
by INTEGRAL separately from the higher temperature blackbody or secondary
steep power law model.
The non-thermal hard component can affect the low energy
spectra ($\lesssim12$\,keV).
The apparent disagreement between the
quiescent emission spectra and 2BB 
for the three AXPs (1E\,2259$+$586, 4U\,0142$+$614 and 1RXS\,J170849.0$-$400910)
might be due to a narrow observational energy band (e.g., $\lesssim$12\,keV).
Therefore, one must not reject 2BB just using the data of the X-ray band.

Although only the upper limit
of a normalization (less than $3.3 \times 10^{-6}$\,photons\,cm$^{-2}$\,s$^{-1}$\,keV
at 30\,keV) 
was reported for AXP\,1E\,2259$+$586 \citep{kuiper2006},
this upper limit
was marginally consistent with a normalization
estimated by our spectral analyses within the error.
The non-detection of the non-thermal hard component above 20\,keV for
AXP\,1E\,2259$+$586 by INTEGRAL \citep{kuiper2006} might imply that
a photon index is very steep and/or there is a spectral cutoff.

To search the non-thermal hard component
for AXP\,1E\,2259$+$586
and also other magnetar candidates would
be very important for understanding the
intrinsic physics of magnetars.
The detailed studies of the non-thermal hard component would be achieved
by simultaneous observations by highly-sensitive detectors
such as the
X-ray imaging spectrometer (0.2-12\,keV; \cite{koyama2007})
and the hard X-ray detector (10-700\,keV; \cite{takahashi2007})
on-board {\it Suzaku} \citep{mitsuda2007}.

\section{Conclusions}
The spectral studies using the
photoelectrically absorbed two blackbody function
(2BB) were presented
for the quiescent emission and the burst of
the magnetar candidates.
The spectra of the quiescent emission were well reproduced by
a 2BB with some exceptions.
The spectra of three AXPs (4U\,0142$+$614, 1RXS\,J170849.0$-$400910 and 1E\,2259$+$586)
seem to have an excess which might be due to a non-thermal hard component discovered by INTEGRAL.
The spectrum of the burst from the SGR candidate SGR\,2013$+$34
was also well reproduced by 2BB.

A strong linear correlations between $kT_{\mathrm{LT}}$ and $kT_{\mathrm{HT}}$
was found using
2BB spectra.
The ratio $kT_{\mathrm{LT}}/kT_{\mathrm{HT}} \sim 0.4$ is almost constant
irrespectively of the objects
and/or emission types (burst
or quiescent emission).
The relationship between $R_{\mathrm{LT}}^2$ and $R_{\mathrm{HT}}^2$ seems
to have a linear correlation.
Considering these correlations,
there seems to be a common emission mechanism among
these objects,
and between the quiescent emission and the burst.
The relationship between the quiescent emission and the burst
might be similar to the relationship between microflares
and an ordinary solar flares of the sun.
The quiescent emission might be due to very frequent small activity
similar to the microflares.
On the other hand, the burst might be due to a relatively
large activity similar to the ordinary solar flare.

\bigskip
We would like to thank an anonymous referee for helpful comments
and suggestions to improve our paper.
This work is based on observations obtained with XMM-Newton,
an ESA science mission with instruments and contributions
directly funded by ESA Member States and NASA.
We would like to thank public data archive of {\it Chandra}.
This research has made use of software provided by the Chandra X-ray Center (CXC)
in the application packages CIAO, ChIPS, and Sherpa.
We acknowledge the use of public data from the Swift data archive.
YEN is supported by the JSPS Research Fellowships for Young Scientists.
This work is supported in part by a special postdoctoral researchers program in RIKEN.


%

\clearpage

\clearpage

\begin{table*}
 \small
 \caption{A summary of magnetar candidates which were employed in our study.}\label{tab:magnetar_list}
 \begin{center}
  \begin{tabular}{lllll}
   \hline\hline
   Object\footnotemark[$*$] & Satellite/Instrument\footnotemark[$\dagger$] & Period\footnotemark[$\ddagger$] & Distance\footnotemark[$\S$] & Ref.\footnotemark[$\|$] \\
   \hline
   SGR\,0526$-$66 & {\it Chandra} ACIS & 2000-2001 & 50 & (1) \\
   SGR\,1627$-$41 & XMM-Newton EPIC, {\it Chandra} ACIS & 2001-2005 & 11 & (2), (3), (4) \\
   SGR\,1806$-$20 & XMM-Newton EPIC, {\it Chandra} ACIS & 2000-2005 & 15\footnotemark[$\#$] & (5), (6) \\
   SGR\,2013$+$34 & {\it Swift} BAT & 2005 & 10 & (7), (8) \\
   SGR\,1819$-$16 & XMM-Newton EPIC & 2003 & 10 & \\
   AXP\,CXOU\,J010043.1$-$721134 & XMM-Newton EPIC, {\it Chandra} ACIS & 2000-2005 & 57 & (9), (10) \\
   AXP\,4U\,0142$+$614 & XMM-Newton EPIC & 2002-2004 & 3 & (11) \\
   AXP\,CXOU\,J164710.2$-$455216 & {\it Swift} BAT, {\it Swift} XRT, {\it Chandra} ACIS & 2005-2007 & 5 & (12), (13), (14), (15) \\
   AXP\,1RXS\,J170849.0$-$400910 & XMM-Newton EPIC & 2003 & 5 & (16), (17) \\
   AXP\,1E\,1841$-$045 & XMM-Newton EPIC & 2002 & 7\footnotemark[$**$] & \\
   AXP\,1E\,2259$+$586 & XMM-Newton EPIC & 2002-2005 & 3 & (18) \\
   \hline
   \multicolumn{5}{@{}l@{}}{\hbox to 0pt{\parbox{180mm}{\footnotesize
       \footnotemark[$*$] Object name of magnetar candidates (SGR\,2013$+$34 denotes SGR candidate SGR/GRB\,050925).
       \par\noindent
       \footnotemark[$\dagger$] Instrument and satellite names from which obtained the data used in our analysis.
       \par\noindent
       \footnotemark[$\ddagger$] Interval which these observations were performed.
       \par\noindent
       \footnotemark[$\S$] Distances to each object in units of kpc (see \cite{woods2006} and references there in).
       \par\noindent
       \footnotemark[$\|$] (1) \citet{kurkarni2003}; (2) \citet{kouveliotou2003}; (3) \citet{wachter2004}; (4) \citet{mereghetti2006b};
                           (5) \citet{kaplan2002}; (6) \citet{mereghetti2005}; (7) \citet{holland2005}; (8) \citet{markwardt2005};
                           (9) \citet{lamb2002}; (10) \citet{majid2004}; (11) \citet{guver2007}; (12) \citet{krimm2006};
                           (13) \citet{campana2006}; (14) \citet{muno2006}; (15) \citet{muno2007}; (16) \citet{oosterbroek2004};
                           (17) \citet{rea2007}; (18) \citet{woods2004}
       \par\noindent
       \footnotemark[$\#$] The latest distance estimate is $d = 8.7$\,kpc \citep{bibby2008}.
       \par\noindent
       \footnotemark[$**$] The latest distance estimate is $d = 8.5$\,kpc \citep{tian2008}.
     }\hss}}
  \end{tabular}
 \end{center}
\end{table*}

\begin{table*}
 \scriptsize
 \caption{XMM-Newton observations of the quiescent emissions of the SGRs.}\label{tab:sgr_obs_xmm}
 \begin{center}
  \begin{tabular}{lllllllllllll}
   \hline\hline
   Object\footnotemark[$*$] & ObsID\footnotemark[$\dagger$] & \multicolumn{2}{l}{Observation Date (MJD)\footnotemark[$\ddagger$]} & \multicolumn{3}{l}{Observation Mode\footnotemark[$\S$]} & \multicolumn{3}{l}{Exposure Time (ks)\footnotemark[$\|$]} & \multicolumn{3}{l}{Source/Background Radii} \\
          &                         & Start & End & pn & MOS1 & MOS2 & pn & MOS1 & MOS2 & pn & MOS1 & MOS2 \\
   \hline
   1627$-$41 & 0204500201 & 53051.590 & 53051.992 & Full & Full & Full & 15.89 & 20.03 & 20.17 & $\timeform{10''}$/$\timeform{10''}$ & $\timeform{10''}$/$\timeform{10''}$ & $\timeform{10''}$/$\timeform{10''}$ \\
   1627$-$41 & 0204500301 & 53252.750 & 53253.131 & Full & Full & Full & 27.06 & 32.00 & 32.10 & $\timeform{10''}$/$\timeform{10''}$ & $\timeform{10''}$/$\timeform{10''}$ & $\timeform{10''}$/$\timeform{10''}$ \\
   1627$-$41 & 0202560101 & 53270.677 & 53271.281 & Small & P-W2 & P-W2 & 26.66 & 36.19 & 49.38 & $\timeform{10''}$/$\timeform{10''}$ & $\timeform{10''}$/$\timeform{10''}$ & $\timeform{10''}$/$\timeform{10''}$ \\
   1806$-$20 & 0148210101 & 52732.566 & 52733.209 & Full & P-W3 & P-W3 & 4.84 & 5.65 & 5.62 & $\timeform{32''}$/$\timeform{32''}$ & $\timeform{28''}$/$\timeform{28''}$ & $\timeform{28''}$/$\timeform{28''}$ \\
   1806$-$20 & 0148210401 & 52919.404 & 52919.663 & Full & P-W3 & P-W3 & 7.61 & 7.07 & 7.23 & $\timeform{32''}$/$\timeform{32''}$ & $\timeform{28''}$/$\timeform{28''}$ & $\timeform{28''}$/$\timeform{28''}$ \\
   1806$-$20 & 0205350101 & 53254.377 & 53254.978 & Small & P-W3 & P-W3 & 30.21 & 39.14 & 39.56 & $\timeform{32''}$/$\timeform{32''}$ & $\timeform{28''}$/$\timeform{28''}$ & $\timeform{28''}$/$\timeform{28''}$ \\
   1806$-$20 & 0164561101 & 53284.706 & 53284.925 & Small & Fast-U & Fast-U & 11.54 & $t$ & $t$ & $\timeform{32''}$/$\timeform{32''}$ & $\cdot\cdot\cdot$ & $\cdot\cdot\cdot$ \\
   1806$-$20 & 0164561301 & 53436.348 & 53436.636 & Small & Fast-U & Full & 7.37 & $t$ & 5.68 & $\timeform{32''}$/$\timeform{32''}$ & $\cdot\cdot\cdot$ & $\timeform{28''}$/$\timeform{28''}$ \\
   1806$-$20 & 0164561401 & 53647.427 & 53647.809 & Small & Fast-U & Full & 22.11 & $t$ & 28.68 & $\timeform{32''}$/$\timeform{32''}$ & $\cdot\cdot\cdot$ & $\timeform{28''}$/$\timeform{28''}$ \\
   2013$+$34 & 0212481201 & 53655.026 & 53655.334 & Full & Full & Full & 22.18 & 25.27 & 25.27 & $\timeform{32''}$/$\timeform{32''}$ & $\timeform{28''}$/$\timeform{28''}$ & $\timeform{28''}$/$\timeform{28''}$ \\
   1819$-$16 & 0152834501 & 52726.191 & 52726.310 & Full & Full & Full & 3.3 & 4.3 & 4.8 & $\timeform{32''}$/$\timeform{80''}$\footnotemark[$\#$] & $\timeform{28''}$/$\timeform{70''}$\footnotemark[$\#$] & $\timeform{28''}$/$\timeform{70''}$\footnotemark[$\#$] \\
   \hline
   \multicolumn{13}{@{}l@{}}{\hbox to 0pt{\parbox{180mm}{\footnotesize
       \footnotemark[$*$] Object name of the SGRs (2013$+$34 denotes SGR candidate SGR/GRB\,050925 and
                          1819$-$16 denotes SGR candidate AX\,J1818.8$-$1559).
       \par\noindent
       \footnotemark[$\dagger$] XMM-Newton observation ID.
       \par\noindent
       \footnotemark[$\ddagger$] Start and end time of observations.
       \par\noindent
       \footnotemark[$\S$] Observation mode for each instrument;
                           full-window mode (Full), small-window mode (Small), partial-w2 mode (P-W2),
                           partial-w3 mode (P-W3) and fast-uncompressed mode (Fast-U).
       \par\noindent
       \footnotemark[$\|$] Net exposure time for each instrument.
                           $t$ denotes the data sets
                           obtained by the MOS cameras in timing mode and not utilized.
       \par\noindent
       \footnotemark[$\#$] The background regions were extracted from an annular region whose center was the source position.
                           The first values are source radii, and the inner radii of the background regions.
                           The second values are outer radii of the background regions.
     }\hss}}
  \end{tabular}
 \end{center}
\end{table*}

\begin{table*}
 \small
 \caption{{\it Chandra} observations of the quiescent emissions of the SGRs.}\label{tab:sgr_obs_cxo}
 \begin{center}
  \begin{tabular}{lllllll}
   \hline\hline
   Object\footnotemark[$*$] & ObsID\footnotemark[$\dagger$] & \multicolumn{2}{l}{Observation Date (MJD)\footnotemark[$\ddagger$]} & Observation Mode\footnotemark[$\S$] & Exposure Time (ks)\footnotemark[$\|$] & Source/Background Radii \\
         &  & Start & End                                &                  &             &       \\
   \hline
   0526$-$66 & 747 & 51547.017 & 51547.539 & FAINT & 39.86 & $\timeform{1''}$/$\timeform{1''}$ \\
   0526$-$66 & 1957 & 52152.937 & 52153.566 & FAINT & 48.45 & $\timeform{1''}$/$\timeform{1''}$ \\
   1627$-$41 & 1981 & 52182.205 & 52182.803 & FAINT  & 48.93 & $\timeform{2''}$/$\timeform{2''}$ \\
   1627$-$41 & 3877 & 52722.169 & 52722.494 & VFAINT & 25.67 & $\timeform{2''}$/$\timeform{2''}$ \\
   \hline
   \multicolumn{7}{@{}l@{}}{\hbox to 0pt{\parbox{180mm}{\footnotesize
       \footnotemark[$*$] SGR names.
       \par\noindent
       \footnotemark[$\dagger$] {\it Chandra} observation ID.
       \par\noindent
       \footnotemark[$\ddagger$] Start and end time of the observations.
       \par\noindent
       \footnotemark[$\S$] FAINT and VFAINT denote the imaging mode, and CC33\_FAINT denotes the timing mode.
       \par\noindent
       \footnotemark[$\|$] Net exposure time.
     }\hss}}
  \end{tabular}
 \end{center}
\end{table*}

\begin{table*}
 \scriptsize
 \caption{XMM-Newton observations of the quiescent emissions of the AXPs.}\label{tab:axp_obs_xmm}
 \begin{center}
  \begin{tabular}{lllllllllllll}
   \hline\hline
   Object\footnotemark[$*$] & ObsID\footnotemark[$\dagger$] & \multicolumn{2}{l}{Observation Date (MJD)\footnotemark[$\ddagger$]} & \multicolumn{3}{l}{Observation Mode\footnotemark[$\S$]} & \multicolumn{3}{l}{Exposure Time (ks)\footnotemark[$\|$]} & \multicolumn{3}{l}{Source/Background Radii} \\
          &                         & Start & End & pn & MOS1 & MOS2 & pn & MOS1 & MOS2 & pn & MOS1 & MOS2 \\
   \hline
   0100$-$721 & 0110000201 & 51834.626 & 51834.867 & E-Full & Full & Full & 20.81 & 14.62 & $g$ & $\timeform{32''}$/$\timeform{32''}$ & $\timeform{28''}$/$\timeform{28''}$ & $\timeform{28''}$/$\timeform{28''}$ \\
   0100$-$721 & 0018540101 & 52233.983 & 52234.303 & Full & Full & Full & 21.16 & 25.73 & $g$ & $\timeform{32''}$/$\timeform{32''}$ & $\timeform{28''}$/$\timeform{28''}$ & $\timeform{28''}$/$\timeform{28''}$ \\
   0142$+$614 & 0112781101 & 52663.920 & 52663.995 & Small & Fast-U & Fast-U & 4.18 & $t$ & $t$ & $\timeform{32''}$/$\timeform{32''}$ & $\cdot\cdot\cdot$ & $\cdot\cdot\cdot$ \\
   1708$-$400 & 0148690101 & 52879.906 & 52880.426 & Full & P-W3 & P-W3 & 26.88 & $p$ & $p$ & $\timeform{20''}$/$\timeform{20''}$ & $\cdot\cdot\cdot$ & $\cdot\cdot\cdot$ \\
   1841$-$045 & 0013340101 & 52552.122 & 52552.192 & Large & Full & Full & 2.34 & 3.54 & 3.57 & $\timeform{12''}$/$\timeform{12''}$ & $\timeform{12''}$/$\timeform{12''}$ & $\timeform{12''}$/$\timeform{12''}$ \\
   1841$-$045 & 0013340201 & 52554.115 & 52554.193 & Large & Full & Full & 4.37 & 6.30 & 6.30 & $\timeform{12''}$/$\timeform{12''}$ & $\timeform{12''}$/$\timeform{12''}$ & $\timeform{12''}$/$\timeform{12''}$ \\
   2259$+$586 & 0038140101 & 52436.378 & 52436.986 & Small & Full & Full & 30.63 & $p$ & $p$ & $\timeform{32''}$/$\timeform{32''}$ & $\cdot\cdot\cdot$ & $\cdot\cdot\cdot$ \\
   2259$+$586 & 0155350301 & 52446.400 & 52446.759 & Small & P-W2 & Full & 17.65 & $p$ & $p$ & $\timeform{32''}$/$\timeform{32''}$ & $\cdot\cdot\cdot$ & $\cdot\cdot\cdot$ \\
   2259$+$586 & 0203550701 & 53579.965 & 53580.030 & Small & P-W2 & Fast-U & 2.66 & $p$ & $t$ & $\timeform{32''}$/$\timeform{32''}$ & $\cdot\cdot\cdot$ & $\cdot\cdot\cdot$ \\
   \hline
   \multicolumn{13}{@{}l@{}}{\hbox to 0pt{\parbox{180mm}{\footnotesize
       \footnotemark[$*$] Object name of the AXPs; CXOU\,J010043.1$-$721134 (0100$-$721), 4U\,0142$+$614 (0142$+$614),
                          1RXS\,J170849.0$-$400910 (1708$-$400), 1E\,1841$-$045 and 1E\,2259$+$586.
       \par\noindent
       \footnotemark[$\dagger$] XMM-Newton observation ID.
       \par\noindent
       \footnotemark[$\ddagger$] Start and end time of the observations.
       \par\noindent
       \footnotemark[$\S$] Observation mode for each instrument;
                           extended full-window mode (E-Full), Full-window mode (Full), small-window mode (Small),
                           fast-uncompressed mode (Fast-U), fast-timing mode (Fast-T), partial-w3 mode (P-W3),
                           large-window mode (Large) and partial-w2 mode (P-W2).
       \par\noindent
       \footnotemark[$\|$] Net exposure time for each instrument.
                           $g$ denotes that the source fell on a gap of the CCD chips,
                           $t$ denotes observations in timing mode, and $p$ denotes that the data
                           sets are affected by a photon pile-up.
                           These data sets were not utilized.
     }\hss}}
  \end{tabular}
 \end{center}
\end{table*}

\begin{table*}
 \small
 \caption{{\it Chandra} observations of the quiescent emissions of the AXPs.}\label{tab:axp_obs_cxo}
 \begin{center}
  \begin{tabular}{lllllll}
   \hline\hline
   Object\footnotemark[$*$] & ObsID\footnotemark[$\dagger$] & \multicolumn{2}{l}{Observation Date (MJD)\footnotemark[$\ddagger$]} & Observation Mode\footnotemark[$\S$] & Exposure Time (ks)\footnotemark[$\|$] & Source/Background Radii \\
         &  & Start & End                                &                  &          &          \\
   \hline
   0100$-$721 & 1881 & 52044.080 & 52045.261 & FAINT  & 98.67 & $\timeform{11''}$/$\timeform{10''}$\footnotemark[$\#$] \\
   0100$-$721 & 4616 & 53031.791 & 53032.009 & VFAINT & 15.56 & $\timeform{2''}$/$\timeform{2''}$ \\
   0100$-$721 & 4617 & 53032.189 & 53032.399 & VFAINT & 15.27 & $\timeform{2''}$/$\timeform{2''}$ \\
   0100$-$721 & 4618 & 53033.904 & 53034.130 & VFAINT & 15.00 & $\timeform{2''}$/$\timeform{2''}$ \\
   0100$-$721 & 4619 & 53042.806 & 53043.023 & VFAINT & 15.04 & $\timeform{2''}$/$\timeform{2''}$ \\
   0100$-$721 & 4620 & 53089.185 & 53089.395 & VFAINT & 15.22 & $\timeform{2''}$/$\timeform{2''}$ \\
   1647$-$455 & 6283 & 53512.860 & 53513.102 & FAINT  & 18.81 & $\timeform{2''}$/$\timeform{2''}$ \\
   1647$-$455 & 5411 & 53539.673 & 53540.141 & FAINT  & 38.47 & $\timeform{2''}$/$\timeform{2''}$ \\
   \hline
   \multicolumn{7}{@{}l@{}}{\hbox to 0pt{\parbox{180mm}{\footnotesize
       \footnotemark[$*$] Object name of the AXPs; CXOU\,J010043.1$-$721134 (0100$-$721) and CXOU\,J164710.2$-$455216 (1647$-$455).
       \par\noindent
       \footnotemark[$\dagger$] {\it Chandra} observation ID.
       \par\noindent
       \footnotemark[$\ddagger$] Start and end time of observations.
       \par\noindent
       \footnotemark[$\S$] FAINT and VFAINT denote the imaging mode.
       \par\noindent
       \footnotemark[$\|$] Net exposure time.
       \par\noindent
       \footnotemark[$\#$] Since the source fell on an off-axis CCD chip, the source region was
                           extracted from an elliptical region with major and minor axes
                           of $\timeform{11''}$ and $\timeform{10''}$, respectively.
     }\hss}}
  \end{tabular}
 \end{center}
\end{table*}

\begin{table*}
 \small
 \caption{{\it Swift} observations of the quiescent emission and the bursts of AXP\,CXOU\,J164710.2$-$455216.}\label{tab:axp_j1647_obs_swift}
 \begin{center}
  \begin{tabular}{lllllllll}
   \hline\hline
   SeqNum\footnotemark[$*$] & \multicolumn{4}{l}{Observation Date (MJD)\footnotemark[$\dagger$]} & \multicolumn{2}{l}{Exposure Time (ks)\footnotemark[$\ddagger$]} & \multicolumn{2}{l}{Source/Background Radii} \\
        & Start (WT) & End (WT) & Start (PC) & End (PC) & WT & PC & WT & PC \\
   \hline
   00230341000\footnotemark[$\S$] & $\cdot\cdot\cdot$ & $\cdot\cdot\cdot$ & $\cdot\cdot\cdot$ & $\cdot\cdot\cdot$ & $\cdot\cdot\cdot$ & $\cdot\cdot\cdot$ & $\cdot\cdot\cdot$ & $\cdot\cdot\cdot$ \\
   00030806001 & 53999.604 & 53999.924 & 53999.604 & 53999.936 & 1.92 & 7.74 & $\timeform{36''}\times\timeform{18''}$/$\timeform{36''}\times\timeform{18''}$\footnotemark[$\|$] & $\timeform{30''}$/$\timeform{30''}$ \\
   00030806002 & 54000.610 & 54000.619 & $\cdot\cdot\cdot$ & $\cdot\cdot\cdot$ & 0.77 & $\cdot\cdot\cdot$ & $\timeform{36''}\times\timeform{18''}$/$\timeform{36''}\times\timeform{18''}$\footnotemark[$\|$] & $\cdot\cdot\cdot$ \\
   00030806003 & 54000.819 & 54001.212 & 54000.819 & 54001.073 & 4.91 & 1.84 & $\timeform{36''}\times\timeform{18''}$/$\timeform{36''}\times\timeform{18''}$\footnotemark[$\|$] & $\timeform{30''}$/$\timeform{30''}$ \\
   00030806004 & 54004.276 & 54004.483 & 54004.343 & 54004.489 & 1.25 & 2.48 & $\timeform{36''}\times\timeform{18''}$/$\timeform{36''}\times\timeform{18''}$\footnotemark[$\|$] & $\timeform{30''}$/$\timeform{30''}$ \\
   00030806006 & 54010.461 & 54010.716 & $\cdot\cdot\cdot$ & $\cdot\cdot\cdot$ & 1.98 & $\cdot\cdot\cdot$ & $\timeform{36''}\times\timeform{18''}$/$\timeform{36''}\times\timeform{18''}$\footnotemark[$\|$] & $\cdot\cdot\cdot$ \\
   00030806007 & 54011.515 & 54011.594 & $\cdot\cdot\cdot$ & $\cdot\cdot\cdot$ & 2.03 & $\cdot\cdot\cdot$ & $\timeform{36''}\times\timeform{18''}$/$\timeform{36''}\times\timeform{18''}$\footnotemark[$\|$] & $\cdot\cdot\cdot$ \\
   00030806008 & 54014.001 & 54014.077 & $\cdot\cdot\cdot$ & $\cdot\cdot\cdot$ & 2.16 & $\cdot\cdot\cdot$ & $\timeform{36''}\times\timeform{18''}$/$\timeform{36''}\times\timeform{18''}$\footnotemark[$\|$] & $\cdot\cdot\cdot$ \\
   00030806009 & 54017.746 & 54017.954 & $\cdot\cdot\cdot$ & $\cdot\cdot\cdot$ & 3.52 & $\cdot\cdot\cdot$ & $\timeform{36''}\times\timeform{18''}$/$\timeform{36''}\times\timeform{18''}$\footnotemark[$\|$] & $\cdot\cdot\cdot$ \\
   00030806010 & 54018.009 & 54018.094 & $\cdot\cdot\cdot$ & $\cdot\cdot\cdot$ & 2.83 & $\cdot\cdot\cdot$ & $\timeform{36''}\times\timeform{18''}$/$\timeform{36''}\times\timeform{18''}$\footnotemark[$\|$] & $\cdot\cdot\cdot$ \\
   00030806011 & 54023.237 & 54023.517 & $\cdot\cdot\cdot$ & $\cdot\cdot\cdot$ & 5.62 & $\cdot\cdot\cdot$ & $\timeform{36''}\times\timeform{18''}$/$\timeform{36''}\times\timeform{18''}$\footnotemark[$\|$] & $\cdot\cdot\cdot$ \\
   00030806012 & 54029.125 & 54029.342 & $\cdot\cdot\cdot$ & $\cdot\cdot\cdot$ & 5.52 & $\cdot\cdot\cdot$ & $\timeform{36''}\times\timeform{18''}$/$\timeform{36''}\times\timeform{18''}$\footnotemark[$\|$] & $\cdot\cdot\cdot$ \\
   00030806013 & 54035.677 & 54035.825 & 54035.678 & 54035.774 & 2.82 & 0.42 & $\timeform{36''}\times\timeform{18''}$/$\timeform{36''}\times\timeform{18''}$\footnotemark[$\|$] & $\timeform{30''}$/$\timeform{30''}$ \\
   00030806014 & 54119.174 & 54119.253 & $\cdot\cdot\cdot$ & $\cdot\cdot\cdot$ & 2.06 & $\cdot\cdot\cdot$ & $\timeform{36''}\times\timeform{18''}$/$\timeform{36''}\times\timeform{18''}$\footnotemark[$\|$] & $\cdot\cdot\cdot$ \\
   00030806015 & 54122.050 & 54122.267 & $\cdot\cdot\cdot$ & $\cdot\cdot\cdot$ & 3.82 & $\cdot\cdot\cdot$ & $\timeform{36''}\times\timeform{18''}$/$\timeform{36''}\times\timeform{18''}$\footnotemark[$\|$] & $\cdot\cdot\cdot$ \\
   \hline
   \multicolumn{9}{@{}l@{}}{\hbox to 0pt{\parbox{180mm}{\footnotesize
       \footnotemark[$*$] {\it Swift} sequence number.
       \par\noindent
       \footnotemark[$\dagger$] Start and end time of the observations for each mode
                                (WT denotes window timing mode, and PC denotes photon counting mode).
       \par\noindent
       \footnotemark[$\ddagger$] Net exposure time for each observation mode.
       \par\noindent
       \footnotemark[$\S$] The observation of a burst.
       \par\noindent
       \footnotemark[$\|$] The source and background regions were extracted from a rectangle region.
                           Two background regions are utilized near both sides of the source region.
     }\hss}}
  \end{tabular}
 \end{center}
\end{table*}

\begin{table*}
 \caption{Spectral parameters of the quiescent emissions of the SGRs observed by XMM-Newton.}\label{tab:sgr_spc_xmm}
 \begin{center}
  \begin{tabular}{lllllllll}
   \hline\hline
   Object\footnotemark[$*$] & ObsID\footnotemark[$\dagger$] & $N_{\mathrm{H}}$\footnotemark[$\ddagger$] & $kT_{\mathrm{LT}}$\footnotemark[$\S$] & $R_{\mathrm{LT}}$\footnotemark[$\|$] &  $kT_{\mathrm{HT}}$\footnotemark[$\S$] & $R_{\mathrm{HT}}$\footnotemark[$\|$] & $F$\footnotemark[$\#$] & $\chi^{2}$ (d.o.f.) \\
          &       & (10$^{22}$ cm$^{-2}$) & (keV)         & (km)         & (keV)         & (km)         &   & \\
   \hline
   1627$-$41 & 0204500201 & 15.98$_{-7.60}^{+15.51}$ & 0.58$_{-0.25}^{+0.35}$ & $<$19.25 & $\cdot\cdot\cdot$ & $\cdot\cdot\cdot$ & $\sim$\,0.05 & 44 (42) \\ 
   1627$-$41 & 0204500301 & 7.53$_{-3.56}^{+6.19}$ & 0.85$_{-0.22}^{+0.29}$ & $<$0.62 & $\cdot\cdot\cdot$ & $\cdot\cdot\cdot$ & 0.08$\pm0.06$ & 29 (65) \\
   1627$-$41 & 0202560101 & 9.00$_{-3.89}^{+6.71}$ & 0.94$_{-0.24}^{+0.31}$ & $<$0.42 & $\cdot\cdot\cdot$ & $\cdot\cdot\cdot$ & 0.06$\pm0.04$ & 54 (47) \\
   1806$-$20 & 0148210101 & 5.18$_{-0.73}^{+0.92}$ & 0.84$_{-0.17}^{+0.23}$ & 1.64$_{-0.53}^{+1.04}$ & 2.62$_{-0.38}^{+0.97}$ & 0.28$_{-0.12}^{+0.10}$ & 11.05$\pm1.92$ & 312 (295) \\
   1806$-$20 & 0148210401 & 5.87$_{-0.55}^{+0.63}$ & 0.85$_{-0.11}^{+0.12}$ & 1.97$_{-0.43}^{+0.65}$ & 3.39$_{-0.58}^{+1.2}$ & 0.20$\pm0.07$ & 12.29$\pm2.4$ & 506 (459) \\
   1806$-$20 & 0205350101 & 5.75$_{-0.19}^{+0.20}$ & 0.96$\pm0.05$ & 2.06$_{-0.17}^{+0.20}$ & 3.19$_{-0.21}^{+0.28}$ & 0.32$\pm0.04$ & 25.43$\pm0.52$ & 2243 (2224) \\
   1806$-$20 & 0164561101 & 5.43$_{-0.35}^{+0.39}$ & 1.02$\pm0.1$ & 1.94$_{-0.27}^{+0.37}$ & 3.92$_{-0.69}^{+1.46}$ & 0.23$\pm0.08$ & 24.65$\pm2.74$ & 737 (741) \\
   1806$-$20 & 0164561301 & 5.64$_{-0.43}^{+0.50}$ & 0.96$\pm0.1$ & 1.98$_{-0.33}^{+0.46}$ & 3.67$_{-0.68}^{+1.54}$ & 0.22$\pm0.08$ & 18.91$\pm4.21$ & 654 (531) \\
   1806$-$20 & 0164561401 & 5.91$_{-0.31}^{+0.34}$ & 0.87$\pm0.06$ & 2.03$_{-0.25}^{+0.31}$ & 3.27$_{-0.34}^{+0.48}$ & 0.22$\pm0.04$ & 13.30$\pm0.5$ & 1026 (1069) \\
   2013$+$34 & 0212481201 & 0.29$_{-0.13}^{+0.15}$ & $\cdot\cdot\cdot$ & $\cdot\cdot\cdot$ & 0.13$_{-0.02}^{+0.02}$ & $<$7.54 & 20.78$\pm20.12$ & 57 (64) \\
   1819$-$16 & 0152834501 & $1.6_{-0.5}^{+0.7}$ & $1.9_{-0.2}^{+0.3}$ & $0.11_{-0.02}^{+0.03}$ & $\cdot\cdot\cdot$ & $\cdot\cdot\cdot$ & 1.3$\pm0.1$ & 59 (74) \\
   \hline
   \multicolumn{9}{@{}l@{}}{\hbox to 0pt{\parbox{180mm}{\footnotesize
       \footnotemark[$*$] Object name of the SGRs (2013$+$34 denotes a SGR candidate SGR/GRB\,050925).
       \par\noindent
       \footnotemark[$\dagger$] XMM-Newton observation ID.
       \par\noindent
       \footnotemark[$\ddagger$] $N_{\mathrm{H}}$ denotes the column density
        with 90 \% confidence level errors.
       \par\noindent
       \footnotemark[$\S$] $kT_{\mathrm{LT}}$ and $kT_{\mathrm{HT}}$ denote the blackbody temperatures with 90 \% confidence level errors.
       \par\noindent
       \footnotemark[$\|$] $R_{\mathrm{LT}}$ and $R_{\mathrm{HT}}$ denote the emission radii with 90 \% confidence level errors.
       \par\noindent
       \footnotemark[$\#$] $F$ denotes a flux in the energy range 2-10 keV in units of $10^{-12}$ ergs cm$^{-2}$ s$^{-1}$ with 68 \% confidence level errors.
     }\hss}}
  \end{tabular}
 \end{center}
\end{table*}

\begin{table*}
 \caption{Spectral parameters of the quiescent emissions of the SGRs observed by {\it Chandra}.}\label{tab:sgr_spc_cxo}
 \begin{center}
  \begin{tabular}{lllllllll}
   \hline\hline
   Object\footnotemark[$*$] & ObsID\footnotemark[$\dagger$] & $N_{\mathrm{H}}$\footnotemark[$\ddagger$] & $kT_{\mathrm{LT}}$\footnotemark[$\S$] & $R_{\mathrm{LT}}$\footnotemark[$\|$] &  $kT_{\mathrm{HT}}$\footnotemark[$\S$] & $R_{\mathrm{HT}}$\footnotemark[$\|$] & $F$\footnotemark[$\#$] & $\chi^{2}$ (d.o.f.) \\
     &  & (10$^{22}$ cm$^{-2}$) & (keV)         & (km)         & (keV)         & (km)         &   & \\
   \hline
   0526$-$66 & 747 & 0.20$_{-0.02}^{+0.03}$ & 0.36$\pm0.03$ & 10.87$_{-1.26}^{+1.66}$ & 0.98$_{-0.14}^{+0.22}$ & 1.06$_{-0.37}^{+0.46}$ & 0.49$\pm0.06$ & 180 (181) \\
   0526$-$66 & 1957 & 0.26$_{-0.04}^{+0.05}$ & 0.30$\pm0.04$ & 15.08$_{-2.87}^{+4.93}$ & 0.70$_{-0.07}^{+0.11}$ & 2.24$_{-0.72}^{+0.85}$ & 0.39$\pm0.06$ & 176 (185) \\
   1627$-$41 & 1981 & 8.47$_{-4.86}^{+6.41}$ & 0.89$_{-0.28}^{+0.57}$ & $<$0.67 & $\cdot\cdot\cdot$ & $\cdot\cdot\cdot$ & $<$0.07 & 36 (34) \\
   1627$-$41 & 3877 & 17.34$_{-7.36}^{+9.89}$ & 0.42$_{-0.13}^{+0.2}$ & 3.56$_{-2.99}^{+33.39}$ & $\cdot\cdot\cdot$ & $\cdot\cdot\cdot$ & $<$0.06 & 22 (21) \\
   \hline
   \multicolumn{9}{@{}l@{}}{\hbox to 0pt{\parbox{180mm}{\footnotesize
       \footnotemark[$*$] Object name of the SGRs.
       \par\noindent
       \footnotemark[$\dagger$] {\it Chandra} observation ID.
       \par\noindent
       \footnotemark[$\ddagger$] $N_{\mathrm{H}}$ denotes the column density with 90 \% confidence level errors.
       \par\noindent
       \footnotemark[$\S$] $kT_{\mathrm{LT}}$ and $kT_{\mathrm{HT}}$ denote the blackbody temperatures with 90 \% confidence level errors.
       \par\noindent
       \footnotemark[$\|$] $R_{\mathrm{LT}}$ and $R_{\mathrm{HT}}$ denote the emission radii with 90 \% confidence level errors.
       \par\noindent
       \footnotemark[$\#$] $F$ denotes the flux in the 2-10\,keV band
                      in units of $10^{-12}$ ergs cm$^{-2}$ s$^{-1}$ with 68 \% confidence level errors.
     }\hss}}
  \end{tabular}
 \end{center}
\end{table*}

\begin{table*}
 \caption{Spectral parameters of a burst of SGR\,2013$+$34 (GRB/SGR\,050925) observed by {\it Swift}.}\label{tab:sgr_050925_spc_swift}
 \begin{center}
  \begin{tabular}{llllllll}
   \hline\hline
   SeqNum\footnotemark[$*$] & $N_{\mathrm{H}}$\footnotemark[$\dagger$] & $kT_{\mathrm{LT}}$\footnotemark[$\ddagger$] & $R_{\mathrm{LT}}$\footnotemark[$\S$] &  $kT_{\mathrm{HT}}$\footnotemark[$\ddagger$] & $R_{\mathrm{HT}}$\footnotemark[$\S$] & $F$\footnotemark[$\|$] & $\chi^{2}$ (d.o.f.) \\
         & (10$^{22}$ cm$^{-2}$) & (keV)         & (km)         & (keV)         & (km)         &   & \\
   \hline
   00156838000 & $\cdot\cdot\cdot$ & 6.6$_{-3.9}^{+5.6}$ & 3.1$_{-1.7}^{+3.7}$ & 20$_{-4}^{+15}$ & $\gtrsim$0.2 & 0.81$\pm0.28$ & 27 (25) \\
   \hline
   \multicolumn{8}{@{}l@{}}{\hbox to 0pt{\parbox{180mm}{\footnotesize
       \footnotemark[$*$] {\it Swift} sequence number.
       \par\noindent
       \footnotemark[$\dagger$] $N_{\mathrm{H}}$ denotes the column density with 90 \% confidence level errors.
       \par\noindent
       \footnotemark[$\ddagger$] $kT_{\mathrm{LT}}$ and $kT_{\mathrm{HT}}$ denote the blackbody temperatures with 90 \% confidence level errors.
       \par\noindent
       \footnotemark[$\S$] $R_{\mathrm{LT}}$ and $R_{\mathrm{HT}}$ denote the emission radii with 90 \% confidence level errors.
       \par\noindent
       \footnotemark[$\|$] $F$ denotes a flux in the energy range 15-150 keV in units of $10^{-6}$ ergs cm$^{-2}$ s$^{-1}$ with 68 \% confidence level errors.
     }\hss}}
  \end{tabular}
 \end{center}
\end{table*}

\begin{table*}
 \small
 \caption{Spectral parameters of the quiescent emissions of the AXPs observed by XMM-Newton.}\label{tab:axp_spc_xmm}
 \begin{center}
  \begin{tabular}{lllllllll}
   \hline\hline
   Object\footnotemark[$*$] & ObsID\footnotemark[$\dagger$] & $N_{\mathrm{H}}$\footnotemark[$\ddagger$] & $kT_{\mathrm{LT}}$\footnotemark[$\S$] & $R_{\mathrm{LT}}$\footnotemark[$\|$] &  $kT_{\mathrm{HT}}$\footnotemark[$\S$] & $R_{\mathrm{HT}}$\footnotemark[$\|$] & $F$\footnotemark[$\#$] & $\chi^{2}$ (d.o.f.) \\
          &       & (10$^{22}$ cm$^{-2}$) & (keV)         & (km)         & (keV)         & (km)         &   & \\
   \hline
   0100$-$721 & 0110000201 & $\lesssim0.06$ & 0.39$\pm0.02$ & 6.94$_{-0.67}^{+1.09}$ & $\cdot\cdot\cdot$ & $\cdot\cdot\cdot$ & 0.09$\pm0.01$ & 264 (275) \\
   0100$-$721 & 0018540101 & $\lesssim0.15$ & 0.29$\pm0.06$ & 11.64$_{-3.35}^{+7.47}$ & 0.64$_{-0.11}^{+0.25}$ & 1.86$_{-1.08}^{+1.17}$ & 0.13$\pm0.09$ & 97 (129) \\
   0142$+$614 & 0112781101 & 0.53$\pm0.01$ & 0.36$\pm0.01$ & 9.38$_{-0.31}^{+0.34}$ & 0.82$_{-0.02}^{+0.03}$ & 0.89$_{-0.08}^{+0.09}$ & 57.26$\pm0.61$ & 1086 (819) \\
   1708$-$400 & 0148690101 & 0.95$\pm0.01$ & 0.48$\pm0.01$ & 4.46$_{-0.10}^{+0.11}$ & 1.49$\pm0.04$ & 0.29$\pm0.02$ & 27.42$_{-0.13}^{+0.08}$ & 1566 (1232) \\
   1841$-$045 & 0013340101 & 1.86$_{-0.13}^{+0.14}$ & 0.52$\pm0.03$ & 3.79$_{-0.46}^{+0.57}$ & 1.99$_{-0.20}^{+0.25}$ & 0.21$_{-0.04}^{+0.05}$ & 17.39$\pm0.83$ & 408 (391) \\
   1841$-$045 & 0013340201 & 1.90$\pm0.11$ & 0.51$\pm0.02$ & 3.97$_{-0.40}^{+0.48}$ & 1.81$_{-0.12}^{+0.14}$ & 0.25$_{-0.03}^{+0.04}$ & 17.16$\pm0.51$ & 583 (641) \\
   2259$+$586 & 0038140101 & $0.59\pm0.01$ & $0.353_{-0.004}^{+0.005}$ & $4.88_{-0.13}^{+0.15}$ & $0.75\pm0.02$ & $0.50_{-0.04}^{+0.05}$ & $12.34\pm0.07$ & 1167 (888) \\
   2259$+$586 & 0155350301 & $0.54\pm0.01$ & $0.390\pm0.006$ & $5.01_{-0.14}^{+0.15}$ & $0.85\pm0.02$ & $0.67_{-0.04}^{+0.05}$ & $33.01\pm0.21$ & 1531 (1053) \\
   2259$+$586 & 0203550701 & $0.59\pm0.03$ & $0.356_{-0.015}^{+0.013}$ & $4.97_{-0.39}^{+0.47}$ & $0.82_{-0.07}^{+0.08}$ & $0.43_{-0.10}^{+0.13}$ & $13.77_{-0.44}^{+0.44}$ & 457 (458) \\
   \hline
   \multicolumn{9}{@{}l@{}}{\hbox to 0pt{\parbox{180mm}{\footnotesize
       \footnotemark[$*$] Object name of the AXPs; CXOU\,J010043.1$-$721134 (0100$-$721), 4U\,0142$+$614 (0142$+$614),
                          1RXS\,J170849.0$-$400910 (1708$-$400), 1E\,1841$-$045 and 1E\,2259$+$586.
       \par\noindent
       \footnotemark[$\dagger$] XMM-Newton observation ID.
       \par\noindent
       \footnotemark[$\ddagger$] $N_{\mathrm{H}}$ denotes the column density with 90 \% confidence level errors.
       \par\noindent
       \footnotemark[$\S$] $kT_{\mathrm{LT}}$ and $kT_{\mathrm{HT}}$ denote the blackbody temperatures with 90 \% confidence level errors.
       \par\noindent
       \footnotemark[$\|$] $R_{\mathrm{LT}}$ and $R_{\mathrm{HT}}$ denote the emission radii with 90 \% confidence level errors.
       \par\noindent
       \footnotemark[$\#$] $F$ denotes a flux in the energy range 2-10 keV in units of $10^{-12}$ ergs cm$^{-2}$ s$^{-1}$ with 68 \% confidence level errors.
     }\hss}}
  \end{tabular}
 \end{center}
\end{table*}

\begin{table*}
 \caption{Spectral parameters of the quiescent emissions of the AXPs observed by {\it Chandra}.}\label{tab:axp_spc_cxo}
 \begin{center}
  \begin{tabular}{lllllllll}
   \hline\hline
   Object\footnotemark[$*$] & ObsID\footnotemark[$\dagger$] & $N_{\mathrm{H}}$\footnotemark[$\ddagger$] & $kT_{\mathrm{LT}}$\footnotemark[$\S$] & $R_{\mathrm{LT}}$\footnotemark[$\|$] &  $kT_{\mathrm{HT}}$\footnotemark[$\S$] & $R_{\mathrm{HT}}$\footnotemark[$\|$] & $F$\footnotemark[$\#$] & $\chi^{2}$ (d.o.f.) \\
          &       & (10$^{22}$ cm$^{-2}$) & (keV)         & (km)         & (keV)         & (km)         &   & \\
   \hline
   0100$-$721 & 1881 & 0.06$_{-0.05}^{+0.06}$ & 0.33$_{-0.04}^{+0.04}$ & 9.65$_{-1.74}^{+2.98}$ & 0.65$_{-0.11}^{+0.23}$ & 1.42$_{-0.85}^{+1.25}$ & 0.12$_{-0.05}^{+0.05}$ & 172 (150) \\
   0100$-$721 & 4616 & $<0.04$ & $\cdot\cdot\cdot$ & $\cdot\cdot\cdot$ & 0.39$\pm0.02$ & 6.84$_{-0.56}^{+1.06}$ & 0.09$\pm0.01$ & 63 (68) \\
   0100$-$721 & 4617 & $<0.41$ & 0.18$_{-0.06}^{+0.25}$ & 21.90$_{-20.07}^{+173.99}$ & 0.44$_{-0.05}^{+21.23}$ & 5.53$_{-5.52}^{+1.24}$ & 0.12$\pm0.04$ & 66 (64) \\
   0100$-$721 & 4618 & $<0.18$ & 0.28$\pm0.1$ & 10.88$_{-3.73}^{+15.89}$ & 0.51$_{-0.09}^{+1.04}$ & 3.02$_{-2.85}^{+2.67}$ & 0.10$\pm0.09$ & 50 (65) \\
   0100$-$721 & 4619 & $<0.04$ & $\cdot\cdot\cdot$ & $\cdot\cdot\cdot$ & 0.41$_{-0.2}^{+0.02}$ & 6.40$_{-0.51}^{+0.88}$ & 0.11$\pm0.01$ & 47 (67) \\
   0100$-$721 & 4620 & $<0.03$ & $\cdot\cdot\cdot$ & $\cdot\cdot\cdot$ & 0.40$_{-0.02}^{+0.01}$ & 6.67$_{-0.41}^{+0.79}$ & 0.10$\pm0.01$ & 70 (68) \\
   1647$-$455 & 6283 & 2.54$_{-0.69}^{+0.81}$ & 0.49$\pm0.06$ & 0.52$_{-0.18}^{+0.31}$ & $\cdot\cdot\cdot$ & $\cdot\cdot\cdot$ & 0.15$\pm0.04$ & 23 (21) \\
   1647$-$455 & 5411 & 1.44$_{-0.28}^{+0.32}$ & 0.58$\pm0.05$ & 0.26$_{-0.05}^{+0.07}$ & $\cdot\cdot\cdot$ & $\cdot\cdot\cdot$ & 0.13$\pm0.01$ & 54 (44) \\
   \hline
   \multicolumn{9}{@{}l@{}}{\hbox to 0pt{\parbox{180mm}{\footnotesize
       \footnotemark[$*$] Object name of the AXPs; CXOU\,J010043.1$-$721134 (0100$-$721) and CXOU\,J164710.2$-$455216 (1647$-$455).
       \par\noindent
       \footnotemark[$\dagger$] {\it Chandra} observation ID.
       \par\noindent
       \footnotemark[$\ddagger$] $N_{\mathrm{H}}$ denotes the column density with 90 \% confidence level errors.
       \par\noindent
       \footnotemark[$\S$] $kT_{\mathrm{LT}}$ and $kT_{\mathrm{HT}}$ denote the blackbody temperatures with 90 \% confidence level errors.
       \par\noindent
       \footnotemark[$\|$] $R_{\mathrm{LT}}$ and $R_{\mathrm{HT}}$ denote the emission radii with 90 \% confidence level errors.
       \par\noindent
       \footnotemark[$\#$] $F$ denotes a flux in the energy range 2-10 keV in units of $10^{-12}$ ergs cm$^{-2}$ s$^{-1}$ with 68 \% confidence level errors.
      }\hss}}
  \end{tabular}
 \end{center}
\end{table*}

\begin{table*}
 \caption{Spectral parameters of the quiescent emission and the bursts of AXP\,CXOU\,J164710.2$-$455216 observed by {\it Swift}.}\label{tab:axp_j1647_spc_swift}
 \begin{center}
  \begin{tabular}{llllllll}
   \hline\hline
   SeqNum\footnotemark[$*$] & $N_{\mathrm{H}}$\footnotemark[$\dagger$] & $kT_{\mathrm{LT}}$\footnotemark[$\ddagger$] & $R_{\mathrm{LT}}$\footnotemark[$\S$] &  $kT_{\mathrm{HT}}$\footnotemark[$\ddagger$] & $R_{\mathrm{HT}}$\footnotemark[$\S$] & $F$\footnotemark[$\|$] & $\chi^{2}$ (d.o.f.) \\
         & (10$^{22}$ cm$^{-2}$) & (keV)         & (km)         & (keV)         & (km)         &   & \\
   \hline
   00230341000\footnotemark[$\#$] & $\cdot\cdot\cdot$ & $\cdot\cdot\cdot$ & $\cdot\cdot\cdot$ & 8.92$_{-1.62}^{+1.34}$ & 1.21$_{-0.35}^{+0.59}$ & 0.35$\pm0.12$ & 8 (12) \\
   00030806001 & 1.72$_{-0.20}^{+0.31}$ & 0.63$_{-0.12}^{+0.07}$ & 2.40$_{-0.44}^{+0.94}$ & 2.08$_{-0.93}^{+38.74}$ & 0.14$_{-0.13}^{+0.36}$ & 27.94$\pm10.25$ & 253 (242) \\
   00030806002 & 1.76$_{-0.55}^{+0.70}$ & 0.73$_{-0.08}^{+0.08}$ & 1.79$_{-0.44}^{+0.7}$ & $\cdot\cdot\cdot$ & $\cdot\cdot\cdot$ & 20.01$\pm4.36$ & 16 (17) \\
   00030806003 & 1.95$_{-0.53}^{+0.86}$ & 0.42$_{-0.15}^{+0.26}$ & 3.03$_{-1.9}^{+9.21}$ & 0.81$_{-0.09}^{+0.35}$ & 1.04$_{-0.81}^{+0.43}$ & 13.98$\pm6.28$ & 101 (92) \\
   00030806004 & 1.79$_{-0.7}^{+0.91}$ & 0.51$_{-0.21}^{+0.28}$ & 2.20$_{-1.66}^{+4.56}$ & 0.93$_{-0.38}^{+3.06}$ & 0.68$_{-0.67}^{+0.5}$ & 13.79$\pm12.1$ & 91 (74) \\
   00030806006 & 0.99$_{-0.32}^{+0.37}$ & 0.75$\pm0.06$ & 1.24$_{-0.23}^{+0.31}$ & $\cdot\cdot\cdot$ & $\cdot\cdot\cdot$ & 11.99$\pm1.2$ & 31 (33) \\
   00030806007 & 1.65$_{-0.39}^{+0.49}$ & 0.67$_{-0.06}^{+0.07}$ & 1.52$_{-0.33}^{+0.48}$ & $\cdot\cdot\cdot$ & $\cdot\cdot\cdot$ & 9.85$\pm1.34$ & 27 (29) \\
   00030806008 & 1.78$_{-0.42}^{+0.55}$ & 0.60$\pm0.05$ & 2.02$_{-0.46}^{+0.69}$ & $\cdot\cdot\cdot$ & $\cdot\cdot\cdot$ & 9.18$\pm1.33$ & 73 (64) \\
   00030806009 & 2.09$_{-0.46}^{+0.57}$ & 0.62$\pm0.06$ & 1.69$_{-0.39}^{+0.58}$ & $\cdot\cdot\cdot$ & $\cdot\cdot\cdot$ & 6.86$\pm1.38$ & 64 (55) \\
   00030806010 & 1.51$_{-0.32}^{+0.4}$ & 0.71$\pm0.06$ & 1.31$_{-0.25}^{+0.34}$ & $\cdot\cdot\cdot$ & $\cdot\cdot\cdot$ & 9.48$\pm0.81$ & 47 (59) \\
   00030806011 & 1.35$_{-0.24}^{+0.28}$ & 0.67$\pm0.04$ & 1.38$_{-0.20}^{+0.26}$ & $\cdot\cdot\cdot$ & $\cdot\cdot\cdot$ & 7.93$\pm0.61$ & 90 (93) \\
   00030806012 & 1.30$_{-0.24}^{+0.29}$ & 0.67$\pm0.05$ & 1.22$_{-0.20}^{+0.25}$ & $\cdot\cdot\cdot$ & $\cdot\cdot\cdot$ & 6.60$\pm0.45$ & 66 (75) \\
   00030806013 & 1.79$_{-0.40}^{+0.49}$ & 0.63$\pm0.06$ & 1.57$_{-0.36}^{+0.54}$ & $\cdot\cdot\cdot$ & $\cdot\cdot\cdot$ & 7.00$\pm0.97$ & 63 (70) \\
   00030806014 & 1.57$_{-0.51}^{+0.72}$ & 0.63$\pm0.09$ & 1.26$_{-0.38}^{+0.68}$ & $\cdot\cdot\cdot$ & $\cdot\cdot\cdot$ & 4.64$\pm1.43$ & 22 (31) \\
   00030806015 & 1.80$_{-0.42}^{+0.53}$ & 0.61$_{-0.06}^{+0.07}$ & 1.38$_{-0.34}^{+0.53}$ & $\cdot\cdot\cdot$ & $\cdot\cdot\cdot$ & 4.77$\pm0.84$ & 49 (47) \\
   \hline
   \multicolumn{8}{@{}l@{}}{\hbox to 0pt{\parbox{180mm}{\footnotesize
       \footnotemark[$*$] {\it Swift} sequence number.
       \par\noindent
       \footnotemark[$\dagger$] $N_{\mathrm{H}}$ denotes the column density with 90 \% confidence level errors.
       \par\noindent
       \footnotemark[$\ddagger$] $kT_{\mathrm{LT}}$ and $kT_{\mathrm{HT}}$ denote the blackbody temperatures with 90 \% confidence level errors.
       \par\noindent
       \footnotemark[$\S$] $R_{\mathrm{LT}}$ and $R_{\mathrm{HT}}$ denote the emission radii with 90 \% confidence level errors.
       \par\noindent
       \footnotemark[$\|$] $F$ denotes fluxes in the energy ranges 15-150 keV in units of $10^{-6}$ ergs cm$^{-2}$ s$^{-1}$ for
                           the burst observation of 00230341000 and 2-10 keV in units of $10^{-12}$ ergs cm$^{-2}$ s$^{-1}$
                           for other observations with 68 \% confidence level errors.
       \par\noindent
       \footnotemark[$\#$] Results for the burst.
      }\hss}}
  \end{tabular}
 \end{center}
\end{table*}

\clearpage


\end{document}